\documentclass[preprint]{aastex}

\usepackage{psfig,emulateapj5}
\bibpunct{(}{)}{;}{a}{}{,}

\newcommand{\chandra}{{\emph{Chandra}}}
\newcommand{\xmm}{\emph{XMM-Newton}}

\newcommand{\kms}{\mbox{\,km\,s$^{-1}$}}
\newcommand{\cmsq}{\mbox{\,cm$^{-2}$}}

\newcommand{\flux}{\mbox{\,erg\,cm$^{-2}$\,s$^{-1}$}}

\newcommand{\fE}{\mbox{\,photon\,cm$^{-2}$\,s$^{-1}$\,keV$^{-1}$}}
\newcommand{\fnu}{\mbox{\,erg\,cm$^{-2}$\,s$^{-1}$\,Hz$^{-1}$}}
\newcommand{\lumin}{\mbox{\,erg\,s$^{-1}$}}
\newcommand{\persec}{\mbox{\,s$^{-1}$}}
\newcommand{\nh}{\mbox{${N}_{\rm H}$}}
\newcommand{\nhi}{\mbox{${N}_{\rm HI}$}}

\newcommand{\aox}{$\alpha_{\rm ox}$}
\newcommand{\auv}{$\alpha_{\rm UV}$}

\newcommand{\CIV}{\ion{C}{4}}

\newcommand{\HI}{\ion{H}{1}}
\newcommand{\delgi}{$\Delta(g-i)$}
\newcommand{\delur}{$\Delta(u-r)$}
\newcommand{\delrz}{$\Delta(r-z)$}
\newcommand{\daox}{$\Delta \alpha_{\rm ox}$}
\newcommand{\HR}{{HR}}
\newcommand{\GHR}{$\Gamma_{\rm HR}$}

\shorttitle{X-RAYING RED QUASARS}
\shortauthors{HALL ET AL.}

\begin{document}

\title{Chandra Observations of Red Sloan Digital Sky Survey Quasars}

\author{
Patrick B. Hall,\altaffilmark{1,2}
S.\ C. Gallagher,\altaffilmark{3}
Gordon T. Richards,\altaffilmark{1,4}
D. M. Alexander,\altaffilmark{5}
Scott F. Anderson,\altaffilmark{6}
Franz Bauer,\altaffilmark{7}
W. N. Brandt,\altaffilmark{8}
Donald P. Schneider\altaffilmark{8}
}
\altaffiltext{1}{Princeton University Observatory, Princeton, NJ 08544.}
\altaffiltext{2}{Department of Physics \& Astronomy, York University, 4700 Keele Street, Toronto, ON M3J 1P3, Canada.}
\altaffiltext{3}{Department of Physics \& Astronomy, University of California, Los Angeles, 475 Portola Plaza, Mail Code 154705, Los Angeles, CA 90095-1547.}
\altaffiltext{4}{Department of Physics \& Astronomy, Johns Hopkins University, 3400 N. Charles St., Baltimore, MD 21218.}
\altaffiltext{5}{Institute of Astronomy, Madingley Road, Cambridge CB3 0HA, UK.}
\altaffiltext{6}{Department of Astronomy, University of Washington, Box 351580, Seattle, WA 98195.}
\altaffiltext{7}{Department of Astronomy, Columbia University, 550 W. 120th Street, New York, NY 10027.}
\altaffiltext{8}{Department of Astronomy \& Astrophysics, The Pennsylvania State University, University Park, PA 16802.}

\begin{abstract}
\noindent{We present short \chandra\ observations of twelve bright ($i$$<$$18$) 
$z_{em}$$\sim$$1.5$ quasars 
from the Sloan Digital Sky Survey 
chosen to have significantly redder optical colors 
than most quasars at the same redshift.  
Of the five quasars with optical properties most consistent with
dust reddening at the quasar redshift, four show indirect evidence of 
moderate X-ray absorption (inferred \nh\,$\sim$\,10$^{22}$\,\cmsq)
with a dust-to-gas ratio $<$1\% of the SMC value.
The remaining seven objects show
no evidence for even moderate intrinsic X-ray absorption.
Thus, while optically red quasars are marginally more likely to show signatures
of X-ray absorption than optically selected quasars with normal colors,
dust-reddened type 1 AGN (as opposed to fully obscured type 2 AGN) 
are unlikely to contribute significantly to the remaining unresolved hard X-ray
background.  
The red quasar population
includes objects with
intrinsically red continua as well as objects with dust-reddened continua.  
Improved sample selection is thus needed to increase our
understanding of either subpopulation.  
To identify dust-reddened quasars likely
to exhibit X-ray absorption, some measure of spectral curvature is
preferable to simple cuts in observed or relative broad-band colors.}
\end{abstract}

\keywords{quasars: general --- X-rays: galaxies --- quasars: absorption lines}

\section{Introduction}

It has long been suspected that optical surveys for
active galactic nuclei are significantly incomplete 
due to orientation-dependent (and perhaps evolution-dependent)
dust obscuration ({Antonucci} 1993).  This suspicion has been confirmed
by recent X-ray and infrared investigations that find higher densities
of active galactic nuclei (AGN) than do optical surveys to comparable
flux limits (e.g., {Ueda} {et~al.} 2003; {Bauer} {et~al.} 2004; {Stern} {et~al.} 2005).  The exact 
fraction of quasars missed by optical surveys is not yet clear, 
as the missing objects appear to be mostly lower-luminosity
AGN (e.g., {Ueda} {et~al.} 2003; {Hao} {et~al.} 2005).  Nonetheless, 
it is clear that a full understanding of quasars and their
contribution to the X-ray background requires proper consideration of
optically obscured quasars.

Indeed, one of the most active topics in X-ray astronomy
today is the origin of the hard X-ray background.  
With the sensitivity of {\em Chandra} and \xmm, $\sim$90\% of the 
X-ray background can be resolved into point sources at $<$6~keV,
but only $\sim$60\% at 6--8~keV (e.g., {Alexander} {et~al.} 2001; {Worsley} {et~al.} 2005).
The spectral shape of the $>10$~keV X-ray background
({Marshall} {et~al.} 1980) clearly indicates that a significant fraction 
of the very hard X-ray background comprises absorbed sources.
The demographics of this absorbed population are uncertain at
present, and it is unclear how many are bona-fide type 2 AGN with
fully obscured broad emission-line and optical continuum regions
as opposed to X-ray-absorbed and optically dust-reddened
but otherwise normal type 1 (broad line)
AGN extincted beyond the flux limits of optical surveys.

Characterizing this
population requires not only statistical 
arguments demonstrating the existence of obscured quasars, but also 
focused investigations targeting the properties of individual
objects that exhibit evidence of obscuration.  Some recent studies
have claimed that quasars with red near-IR colors
($J-K>2$) from the Two-Micron All-Sky Survey (2MASS) represent
a population of previously undiscovered quasars that could partially
account for much of the ``missing'' hard X-ray background sources
({Cutri} {et~al.} 2002; {Wilkes} {et~al.} 2002, 2005).  In this paper we take a complementary
approach and investigate the nature of quasars that have
{\em relative} UV/optical colors that are redder than average (at a given 
redshift).  This selection has the benefit of being less sensitive to color
differences due to emission lines moving in and out of the bands and of
exhibiting larger differences due to extinction, since the effects of
extinction are strongest at UV wavelengths.  We use {\em Chandra} observations
of these quasars to look for additional signs of extinction,
such as a harder X-ray spectrum and lower than expected X-ray flux.

We present the construction of our sample in \S~2,
our X-ray observations and data analysis in \S~3,
the results of our observations in \S~4,
a discussion of our results in \S~5,
and our conclusions in \S~6.
We adopt $H_{\rm 0}=70$~km\,s$^{-1}$\,Mpc$^{-1}$, $\Omega_{\Lambda}=0.7$,
and $\Omega_m=0.3$ (e.g., {Spergel} {et~al.} 2006).

\section{Sample Selection}
\label{sec:sample}

\subsection{Why Use Relative Optical Colors?}

One of the primary problems with color-selecting a sample of ``red'' quasars
for further study is that quasar colors are a strong function of redshift.
Large-area digital surveys such as the Sloan Digital Sky Survey (SDSS;
{York} {et~al.} 2000) are now enabling astronomers to characterize the
quasar color distribution as a function of redshift ({Richards et al.} 2001) and
are, in fact, \emph{resolving} their color distribution ({Richards} {et~al.} 2003).
While a fixed observed-frame color cut such as $B-K>5$ does select
quasars that are {\em apparently} red, as discussed by {Richards} {et~al.} (2003)
this approach fails to distinguish whether
the quasar is intrinsically red (i.e., has a red [optically steep]
power-law spectrum), has been reddened by dust extinction,
or appears red because of strong emission or absorption in one of the bands  
or long-wavelength host galaxy contamination.  For quasars of known redshift, 
it is possible to use SDSS colors alone to account for the redshift dependence
of quasar colors and, for sufficiently large reddening, to distinguish between
a steeper power-law continuum and the curvature of a dust-reddened continuum.

Thus, to define a more redshift-independent red quasar sample, we take advantage
of the well-defined structure in the color-redshift relation (see Fig.~1).
For each of our quasars we can compute a {\em relative} color by
subtracting the median color of quasars at that redshift ({Richards et al.} 2001).
Relative colors are thus more nearly equally sensitive to reddened quasars
at all redshifts and can be used to classify quasars as a function of color
alone.  For example, a quasar that has a relative $u-g$ color of 0.2
($\Delta(u-g)=0.2$) is redder than the average quasar by 0.2 magnitudes
in observed $u-g$ regardless of its redshift.

Such selection should yield a more homogeneous red AGN sample than a
fixed observed-frame color cut such as $J-K>2$ ({Cutri} {et~al.} 2002).
The relative color \delgi\ probes the shape of the UV/optical continuum
due to the accretion disk at most redshifts studied by the SDSS,
whereas a $J-K$ selected sample
measures the shape of the near-IR bump for $z\lesssim0.25$
and the ratio of the near-IR to optical flux densities for $0.25<z<1$.
Furthermore, although a $J-K>2$ color cut does include dusty quasars,
it also selects only slightly redder than average quasars with $z<0.5$
(see Fig.~5 of {Barkhouse} \& {Hall} 2001 and \S\,6.2 of {Maddox} \& {Hewett} 2006).  

Near-IR selection of AGN is superior to optical selection when one is
trying to {\em avoid} incompleteness caused by dust extincting objects
below the survey flux limit. 
Yet, specifically {\em because} the UV/optical colors suffer more
extinction, UV/optical colors are better able to {\em distinguish} 
between dust-reddened quasars and
unreddened quasars with steeper than average continua.
Our SDSS-based sample has one additional advantage in that 
the flux limit for the optically 
selected red quasars was determined in the $i$-band instead
of the $B$-band, resulting in less of a bias against dust-reddened
quasars (as compared to most other optical
surveys).  Thus an investigation of red quasars using relative optical
colors of SDSS quasars provides a happy medium for exploring the red
quasar population: it has the benefits of relatively long wavelength
selection, but covers wavelengths considerably affected by dust
reddening.

\subsection{Selection Details}

\begin{figure}
\epsscale{1.0}
\plotone{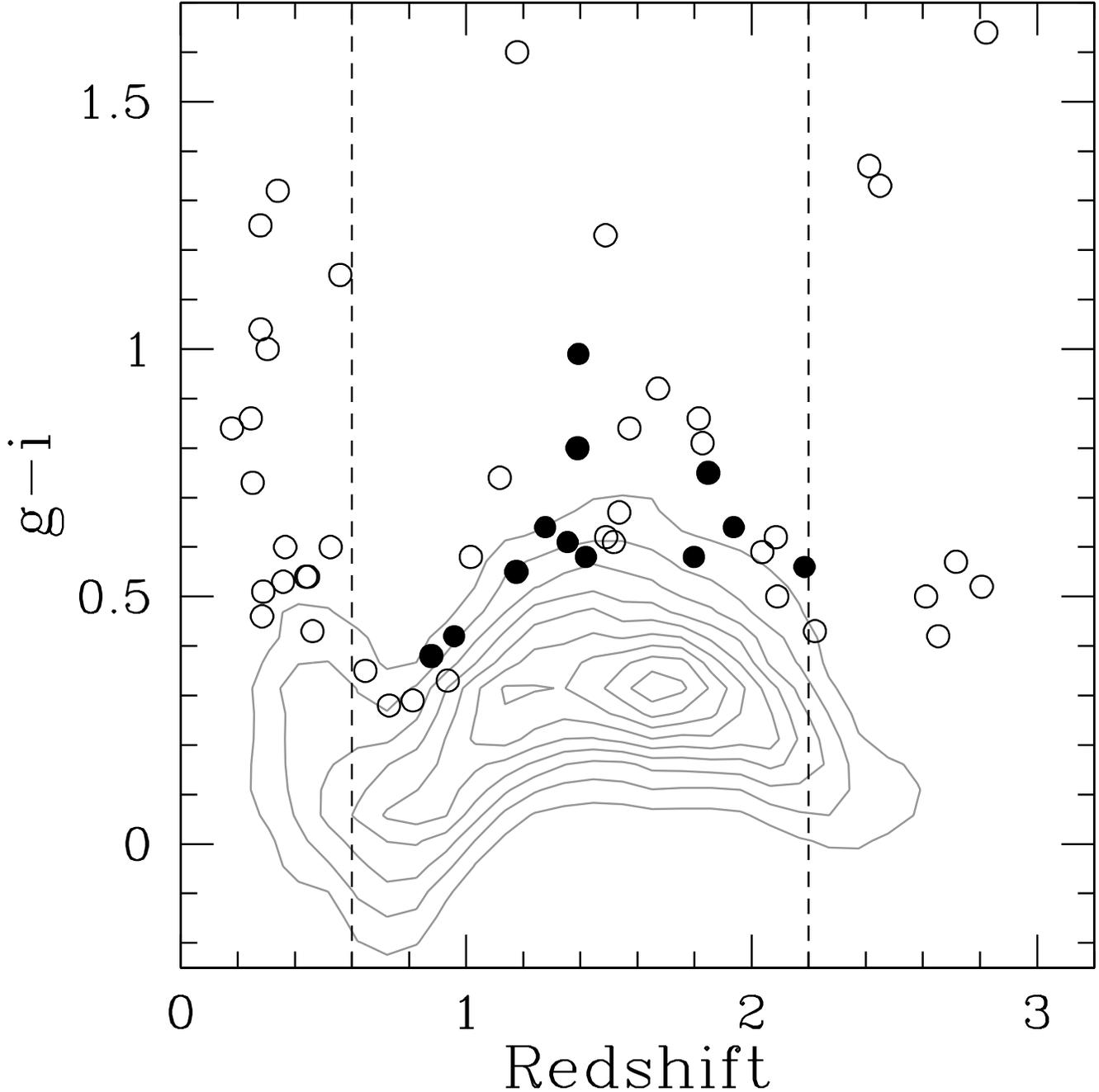}
\caption{The $g-i$ colors of the SDSS EDR quasar sample from which
our targets were chosen.  The contours
show the distribution of $g-i$ colors for all 3814 quasars from the EDR.
Circles denote optically selected red quasars with $\Delta(g-i)>0.2$.
The filled black circles are the 16 red quasars that met our other selection
criteria: $0.6\leq z \leq 2.2$ (vertical dashed lines), $i<18$, no broad
absorption lines (BALs), radio-quiet, and low Galactic $N_H$.  
(Of these, the 12 listed in Table~1 were awarded \chandra\ time.)
Only $\sim$11.5\% of non-BAL, $i<18$ quasars in our redshift range
had relative colors red enough to be considered as a candidate red quasar, but
many of the reddest quasars are contaminated by host galaxy light (at $z<0.6$)
or are BAL quasars, either of which was grounds for exclusion from our sample.
\label{fig:relgi}
}
\end{figure}

Our initial sample was selected from the 3814 Early Data Release 
(EDR) quasars ({Schneider} {et~al.} 2002), giving priority to bright objects 
(dereddened $i<18.0$) with small Galactic \nh.  We considered only
spatially unresolved quasars with \mbox{$0.6\le z_{\rm em}\le 2.2$}
and \delgi\,$>0.2$; see Figure~\ref{fig:relgi}.  The color cut selects 
the reddest $\sim$11.5\% of non-BAL, $i<18$ quasars at each redshift,
and the redshift selection avoids
host galaxy contamination and reduces the effect of
absorption from the Ly$\alpha$ forest.  Avoiding resolved sources also
reduces the chances that the red colors are due to host galaxy
contamination.  The sample was vetted to exclude 
quasars with broad absorption line (BAL) troughs ({Reichard} {et~al.} 2003b) which are
known to cause both optical reddening and X-ray absorption.
(e.g., {Gallagher} {et~al.} 2006).  This was not possible for the eight
$z_{\rm em}<1.6$ quasars where \ion{C}{4} $\lambda$1549, 
the transition that most often exhibits a BAL trough,
is not in the SDSS spectroscopic bandpass.  Based on a raw high-ionization
SDSS BAL fraction of 9.85\% (Trump et al. 2006), we estimate there is
a 38\% chance that exactly one of these eight quasars is a high-ionization BAL
quasar, a 15\% chance that exactly two are, and a 3\% chance that three or more are.
We also excluded quasars detected in the 20~cm FIRST survey ({Becker}, {White}, \& {Helfand} 1995),
since radio-powerful quasars have different X-ray properties (e.g., {Shastri} {et~al.} 1993).  
Table~\ref{tab:opt} lists the 12 targets for which we obtained
X-ray data.  Figure~\ref{fig:optspec1} shows their SDSS spectra.

An obvious question to ask is whether our selection criterion recovers
quasars selected by $J-K>2$ (and vice versa).  
In Figure~\ref{fig:relj-k} we show the relative $g-i$ color versus the
relative $J-K$ color for SDSS-DR3 quasars also detected by 2MASS
({Schneider} {et~al.} 2005).  
Objects consistent with being dust reddened form a tail
with red $\Delta(g-i)$ colors but undistinguished $\Delta(J-K)$ colors.  
In terms of observed $J-K$ color we find that while our \delgi$\;>0.2$ criterion
does indeed recover some quasars with $J-K>2$, only 1/3 of EDR quasars with
$J-K>2$ also have \delgi$\;>0.2$.
Since the tail of red \delgi\ colors comes predominantly from dust
reddening ({Hopkins} {et~al.} 2004), $J-K>2$ selection must miss some dust-reddened
quasars and include some that are not dust reddened.
For example, none of our twelve targets have $J-K>2$.\footnote{For four of our
targets without 2MASS $J$ or $K$ detections, we performed aperture photometry
on calibrated 2MASS Atlas images.}
We discuss $J-K>2$ selection further in \S\,\ref{sec:jk2}.

Note that our selection criteria were designed to select quasars that
suffer from extinction (and reddening) by dust grains, not ones that
simply have intrinsically redder than average power-law continua.
Unfortunately, the range of power-law slopes in quasars is such 
that it is not possible to do this with complete certainty.
However, any unreddened but intrinsically red quasars 
that `contaminate' our sample are still of interest,
since the relationship between the X-ray properties of quasars and their
optical/UV colors is not well characterized.  

\subsection{Damped Ly$\alpha$ Absorbers}	\label{sec:dla}

In our selection of targets, we did not reject objects with strong
intervening absorption lines in their optical spectra.  At first
glance not rejecting such systems seems reasonable given the amount of
reddening expected from a typical intervening absorption system.  As
shown by {Richards} {et~al.} (2003) and {Hopkins} {et~al.} (2004), in a statistical sense
the reddening observed in SDSS quasars is dominated by SMC-like dust
at the quasar redshifts.
However, quasars with intervening absorption lines are three times as
likely to have highly reddened spectra as quasars without them ({York} {et~al.} 2006).
And since SMC-like dust has a nearly featureless, 
nearly power-law exinction curve, it is not always possible
{\em for a specific quasar} to determine whether
the dust is intrinsic to the quasar or along the line of sight.

We can estimate the amount of optical reddening that is expected under
the hypothesis that the galaxies causing absorption lines in some of
our quasars are the source of reddening.  Following Rao \& Turnshek
(2000, Fig. 25),\nocite{rt00} if the rest-frame equivalent widths are
$EW_{\rm MgII 2796}>0.5$\,\AA\ and $EW_{\rm FeII 2600}>0.5$\,\AA, there
is a $50\%$ chance the \ion{Mg}{2} system comes from a DLA with $N_{\rm
HI}>2\times10^{20}$\cmsq, which gives $E(B-V)>0.005$ using the SMC
extinction law.  Quasars with such candidate DLAs are noted in
Table~\ref{tab:opt}.  The two which also show \ion{Ca}{2}
absorption lines, J1310+0108 and J1323$-$0021,
will be specially marked as they are the most
likely to be affected by intervening extinction ({Wild} \& {Hewett} 2005).

Note that for the {\em maximum} $N_{\rm HI}$ found for {\em any} 
DLA along a quasar sightline to date, $5\times10^{21}$\cmsq, 
the SMC extinction would be only $E(B-V)=0.11$.  
However, given recent work demonstrating that intervening absorbers
sometimes {\em can} redden quasars ({Khare} {et~al.} 2005), it may be that 
these {\em particular} systems have LMC or MW extinction laws which
would require a multiplication of $E(B-V)$ by factors of 2 or 9,
respectively.  If that is the case, then the redness of such
objects may not be intrinsic and they cannot be used to address
the issue of the X-ray properties of {\em intrinsically} red or 
reddened quasars.  
We consider this issue further in \S\,\ref{sec:notes}.

\section{X-ray Observations and Data Analysis}	\label{sec:data}

\chandra\ observed our 12 red quasar targets between 2003 November 09
and 2004 September 29 (see Table~\ref{tab:log}).  
Each target was observed at the aimpoint of the back-illuminated S3 
CCD of the Advanced CCD Imaging Spectrometer (ACIS; {Garmire} {et~al.} 2003)
in faint mode.

In general, data analysis followed the procedure detailed in 
{Gallagher} {et~al.} (2005), which we outline briefly. Both aperture photometry
and the CIAO version 3.2 (see http://cxc.harvard.edu/ciao)
wavelet detection tool {\em wavdetect} ({Freeman} {et~al.} 2002) were used in
the soft (0.5--2~keV), hard (2--8~keV), and full (0.5--8~keV)
bands to determine the measured counts for a point source in each band.  
Except for J1133$+$0058, the quasars were each significantly detected
with full-band counts ranging from 7 to 237; the difference in counts
between {\em wavdetect} and aperture photometry was always
$\le3$~counts.  The background 
was in all cases negligible ($<$1 count in the source region). 
At the median flux of our sample 
the source density is 8$\times10^{-7}$ arcsec$^{-2}$ (e.g., Bauer {et~al.} 2004),
so the chance of a misidentification is extremely small.

For each target, to provide a coarse quantitative measure of the
spectral shape we calculated the hardness ratio 
\HR=$(h$$-$$s)/(h$$+$$s)$, where $h$ and $s$ refer to the hard- and soft-band
counts, respectively.  A typical, radio-quiet quasar has a power-law
continuum in the 0.5--10.0~keV band characterized by the photon index,
$\Gamma$, and the 1~keV normalization, $N_{\rm 1 keV}$: $f_{\rm
E}=N_{\rm 1 keV}E^{-\Gamma}$ \fE.  
From X-ray spectroscopic studies,
$\Gamma$ is found to average $2.0\pm0.25$ for radio-quiet
quasars (e.g., {George} {et~al.} 2000; {Reeves} \& {Turner} 2000); such an average quasar
would be observed to have \HR\,$\approx-0.63$.

To transform the observed \HR\ into $\Gamma$, the X-ray spectral
modeling tool {\em XSPEC} ({Arnaud} 1996) was used.
The detector response to incident power-law spectra with varying $\Gamma$ was 
simulated.  The
hardness ratio and errors were compared to the modeled \HR\ to
determine the \GHR\ that would generate the observed \HR.  The
uncertainties on \GHR\ reflect the statistical uncertainties in the 
\HR.  The modeled full-band count rate was normalized to the observed
full-band count rate to obtain the power-law normalization, $N_{\rm 1
keV}$.  Using $N_{\rm 1 keV}$ and \GHR, the 0.5--8.0~keV flux, $F_{\rm
X}$, and the flux density at rest-frame 2~keV, $f_{\rm 2 keV}$, were
calculated.  The errors quoted for these two values are the Poisson
errors ({Gehrels} 1986) from the full-band counts.

Lastly, we calculated the optical/UV to X-ray index
\aox=$\log(f_{\rm 2 keV}/f_{\rm 2500})/\log(\nu_{\rm 2 keV}/\nu_{\rm 2500})$
and its uncertainty.  
The quantity $f_{\rm 2500}$ is the average
flux density within the rest-frame range $2500\pm25$\,\AA\ in the
SDSS spectrum, scaled to account for flux outside the spectroscopic
fiber aperture.  We took the scale factor to be the flux ratio
between the object's photometric PSF magnitude and
its fiber magnitude as synthesized from the spectrum,
with both magnitudes measured in the filter into which
rest-frame 2500\,\AA\ is redshifted.  
As detailed in \S2.2 of {Gallagher} {et~al.} 2005,
the quoted errors in \aox\ incorporate both the
Poisson uncertainty in $f_{\rm 2keV}$ and the estimated effects of UV
variability between the SDSS and {\em Chandra} observations ({Ivezi{\' c}} {et~al.} 2004),
which are on average 500 days apart in the quasar rest frame.
We also calculated \daox $=$ \aox $- \alpha_{\rm ox} (L_{2500})$; a
quasar's \daox\ is its \aox\ relative to the expected \aox\ for a
quasar of its luminosity ({Strateva} {et~al.} 2005, Eq. 6; see also
{Steffen} {et~al.} 2006).  These
quantities are presented in Table~\ref{tab:xcalc}.

\subsection{X-ray Spectral Fitting}
\label{sec:xspecfit}

For those six quasars with $>50$ full-band counts, we also performed X-ray
spectral fitting.  The spectra were extracted using
2$\farcs$5-radius source cells.
Our spectral model was a simple absorbed power-law continuum with both
Galactic and intrinsic absorption by neutral gas with solar abundances
at the quasar redshift; the photon statistics do not warrant a
more complex model.  In this low-count regime, the spectra are not
binned and the fit is performed with {\em XSPEC} by minimizing the
$C$-statistic ({Cash} 1979).  The best-fitting $\Gamma$ and intrinsic
\nh\ for the six quasars with $>50$ counts are listed in
Table~\ref{tab:xspec}.  
{
Since these quasars have the highest X-ray fluxes in our sample, 
they are by definition unlikely to be absorbed in the X-ray; indeed,
for only one of them is there a significant detection of absorption.}
This object, J1323$-$0021, exhibits the most metal-rich intervening
absorption system currently known (\S\ref{1323}).

\section{Results} \label{sec:res}
\subsection{Correlations Between X-ray and Optical Properties} \label{sec:xo}

We begin our X-ray analysis by searching for evidence of
X-ray absorption that may occur in conjunction with optical
dust extinction and reddening.  The signatures of absorption
in snapshot X-ray observations are flatter than average 
X-ray photon indices (i.e., $\Gamma \lesssim 1.5$),
since absorption preferentially removes softer X-rays,
and unusual values of the optical to X-ray flux ratio, \aox, or the
luminosity-corrected optical to X-ray flux ratio, \daox.  

We first examine the relationship between \GHR\ and \delgi, which is
shown in Figure~\ref{fig:ghrdelg}.\notetoeditor{Figure 4 should be typeset
with the two panels of the figure side by side across the FULL top or bottom
of a page, as wide as possible.}  {Richards} {et~al.} (2003) showed that most
quasars have \delgi\ between $-0.2$ and $0.2$ and that there exists a
tail of redder quasars consistent with being dust reddened.  The typical
range of $\Gamma$ is found to be $2.0\pm0.25$ for radio-quiet quasars
(e.g., {George} {et~al.} 2000; {Reeves} \& {Turner} 2000).  Dust causes quasars to appear
redder in the optical, while gas absorbs soft X-ray photons.  
For any observed dust reddening, the accompanying X-ray absorption will
be greater (and thus the \GHR\ lower) if the dust-to-gas ratio is smaller.
Quasars in the lower right hand corner of panel (a) 
in Figure~\ref{fig:ghrdelg} are potentially obscured quasars.  

In panel (b) of Figure~\ref{fig:ghrdelg} we show \GHR\ vs. the
ultraviolet spectral slope, \auv\ ($F_\nu\propto\nu^{\alpha_{\rm UV}}$).
\auv\ measures the quasar's underlying
continuum slope at fixed rest-frame wavelengths that are relatively
devoid of emission line flux and is thus a more precise measurement
of the continuum color than \delgi.
Our new data appear to reaffirm the correlation between \GHR\ and
\auv\ found by {Gallagher} {et~al.} (2005). 
Combining both datasets, the chance of no correlation is only 
0.03\% using Kendall's $\tau$ test.  
Even considering only quasars with \daox~$\geq -0.1$,
the chance of no correlation is still only 0.95\%.
In the plot we show the BCES(Y$|$X) 
({Akritas} \& {Bershady} 1996) linear fit to the full combined datasets,
$\Gamma_{\rm HR}=(0.54\pm0.15)\alpha_{UV}+(2.07\pm0.13)$.
More work, however, is needed to disentangle a correlation between
\GHR\ and intrinsic UV spectral slope from a correlation induced
by the effects of dusty gas on both parameters.
That is, an intrinsically red quasar might have an intrinsically
hard X-ray spectrum, whereas a dust-reddened quasar might have
an intrinsically soft X-ray spectrum that appears hard due to
X-ray absorption.

We next search for for any trends between color and \daox\, 
as shown in Figure~\ref{fig:daoxdelg}.\notetoeditor{Figure 5 should be
typeset with the two panels of the figure side by side across the FULL 
top or bottom of a page, as wide as possible.}  In panel (a) the lines
show the expected relationship between \daox\ and \delgi\ for
neutral gas with three different SMC-like dust-to-gas ratios.  
For an SMC dust-to-gas ratio, even minimal X-ray absorption
can be accompanied by heavy extinction of optical light by dust,
and the ratio of emergent X-ray to optical fluxes
(in other words, \daox) increases.  
But as the dust-to-gas ratio decreases, X-ray wavelengths become
increasingly affected relative to the optical.
The open points are roughly consistent with a typical 
SMC reddening as shown by the solid line.  
The filled points, however, may require a different explanation.
Both panels of Figure~\ref{fig:daoxdelg}
show that there is intrinsic scatter in \daox\ which is
independent of a quasar's optical/UV color, but that quasars 
with relatively red colors may be more likely to be X-ray faint
than are quasars with relatively blue colors.\footnote{The converse,
that relatively X-ray faint quasars 
are more likely to be abnormally red, is also probably true.}
When \auv\ is used as a measure of optical color
to split the full plotted sample in half at \auv=$-$0.75,
an invariant Kolmogorov-Smirnov (Kuiper) test shows that
the \daox\ distributions for the two halves are different
at the 89.2\% significance level.
However, when the sample is split in half at \delgi=0.26,
there is only a 55.8\% chance the distributions are different.

To test whether the extrema in these figures are consistent with
absorption in the X-ray, in Figure~\ref{fig:ghrdaox} we show \GHR\
vs.\ \daox\ overlaid on curves that demonstrate the effects of X-ray
absorption on a quasar of intrinsic \GHR=2 and \daox=0.
The heavy curves are for gas with 
various dust-to-gas ratios at a fixed $z=1.5$, 
while the dotted curves are for completely dust-free neutral gas
at the redshifts shown.
Dust-free gas absorbs soft X-rays, but does not affect the
optical spectrum until the absorption becomes Compton-thick.
Thus, the effect of such gas in Figure~\ref{fig:ghrdaox} 
is to decrease both \daox\ and \GHR.
The three points closest to the lower left-hand corner appear consistent
with the absorption hypothesis and inconsistent with being due to
intrinsic scatter in \GHR\ and \daox.  Furthermore, the distribution
of points in the diagram is not consistent with random scatter around
\GHR=2 and \daox=0, but is consistent with quasars with intrinsic \GHR=2
and \daox=0 being affected by absorption of varying dust-to-gas ratios,
as indicated by the various tracks, HR=2 and \daox=0.  However, with the
typical measurement accuracy and probable intrinsic scatter of \GHR\ and
\daox, this Figure by itself can only be used to unambiguously identify 
moderately reddened quasars if they have {\em low} dust-to-gas ratios.
Moderate reddening with a high dust-to-gas ratio can produce a
marked effect on a quasar's optical spectrum without substantially
affecting its X-ray properties;  a quasar's optical/UV spectrum is 
an invaluable source of information on the reddening of such quasars.
(See \S\,\ref{sec:notes} and \S\,\ref{sec:dis} for specific examples.)

\subsection{Notes on Individual Quasars}	\label{sec:notes}
\begin{figure}
\epsscale{1.0}
\plotone{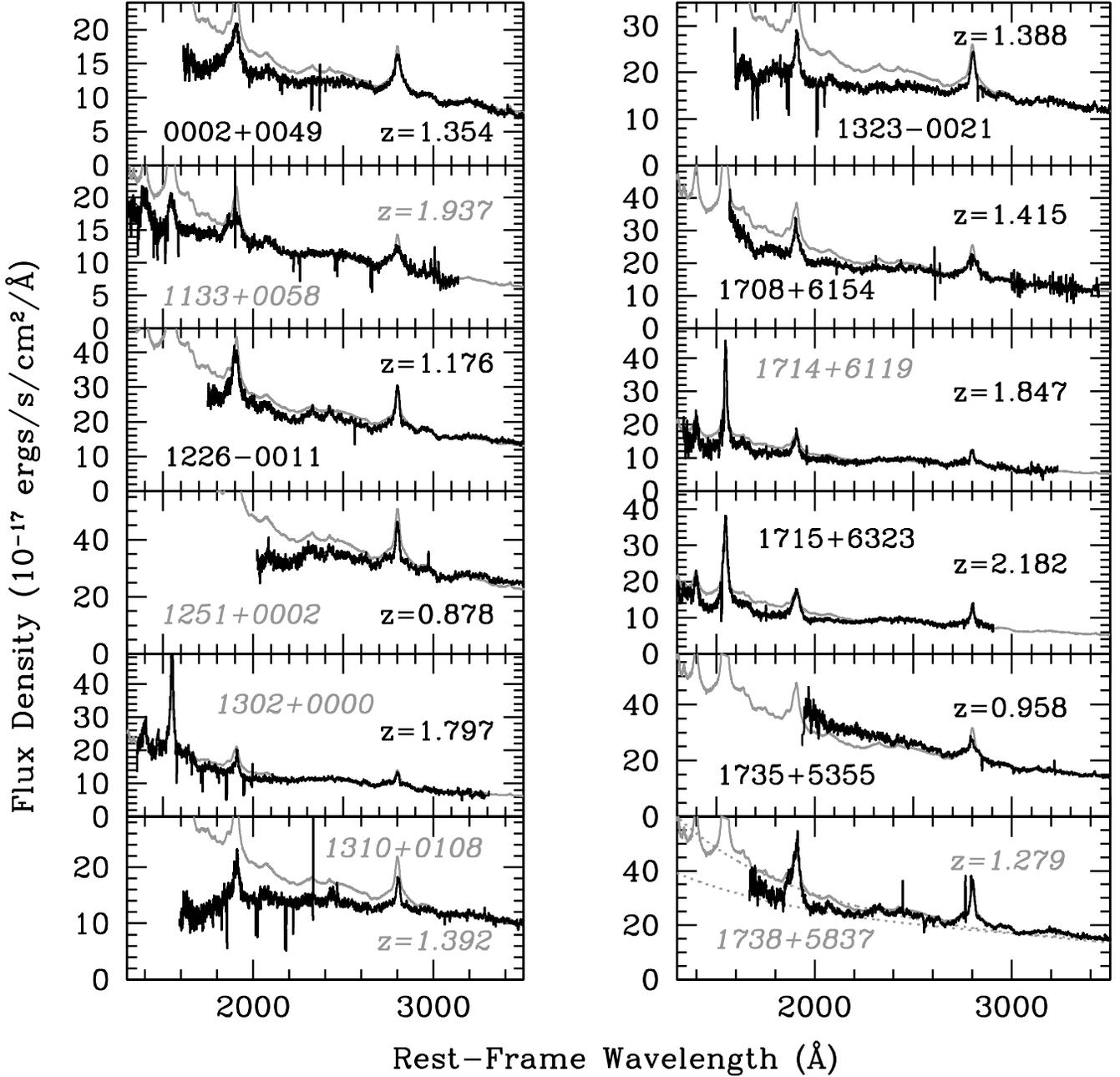}
\caption{SDSS optical/UV spectra of our red quasar candidates.  The
gray curves in each panel show the {Vanden Berk} {et~al.} (2001) composite quasar
spectrum for reference.  
To further illustrate the difference between normal and red quasars,
the dotted gray curves in the bottom right panel 
show power-law continua with $\alpha_\nu=-0.5$ (normal) and $-1.0$ (red).
Quasars with negative values of \daox\ have their names shown in gray italics.
The three of those quasars which have the hardest \GHR\ values 
also have their redshifts shown in gray italics.
\label{fig:optspec1}
}
\end{figure}

Before we can explore the X-ray properties of dust-reddened quasars as
a population, we must examine the diagnostic information in
Figures \ref{fig:optspec1}, \ref{fig:ghrdelg}, and \ref{fig:daoxdelg}
to determine which objects in our sample are most likely to be
reddened by dust versus intrinsically red.  Both of these groups could
suffer from X-ray absorption, which does not always occur in
conjunction with dust extinction (e.g. {Brandt}, {Laor}, \& {Wills} 2000).  The shape of
the optical/UV continuum, the relationship between the relative
colors, and the X-ray properties affect this determination.  We now examine
these parameters for each of our objects.

\subsubsection{J0002+0049; $z_{\rm em}=1.355$, $\Delta(g-i)=0.396$, $\alpha_{\rm UV}=-0.97$, $\Gamma_{\rm HR}=2.13$, $\Delta\alpha_{\rm ox}=0.11$} \label{0002}

The optical spectrum shows moderate reddening, for which the intervening
absorption system is likely too weak to be responsible.  
This quasar has the softest X-ray spectrum in the sample and \daox\ is 
positive, thus there is no sign of X-ray absorption; the results from
spectral-fitting are consistent with this conclusion.  
These properties might lead us to believe that this quasar is 
intrinsically red.  However, there is evidence for curvature in the optical
spectral energy distribution (SED) from the SDSS photometry.  
Specifically \delur\ is larger than \delgi, which is larger than \delrz,
as expected for a dust-reddened object.  (For an object with a power-law
continuum and unexceptional emission-line properties \delur, \delgi, and
\delrz\ would be equal within the uncertainties.)  Thus we conclude that
the object is reddened by SMC dust with a normal dust-to-gas ratio;
this leads to noticable curvature of the optical spectrum but
negligibly effects the X-ray spectrum.

\subsubsection{J1133+0058; $z_{\rm em}=1.937$, $\Delta(g-i)=0.382$, $\alpha_{\rm UV}=-1.15$, $\Gamma_{\rm HR}$ {\rm unknown}, $\Delta\alpha_{\rm ox}=<-0.56$} \label{1133}

The optical spectrum is consistent with mild dust reddening rather than a
flatter than typical power-law slope.  Dust reddening is also indicated by the
fact that \delur\ $>$ \delgi\ $>$ \delrz.  There is a strong intervening
absorption system (a probable DLA) at $z=1.786$.
This quasar was undetected in our \chandra\ observation
and must thus be very heavily absorbed in the X-ray 
(equivalent \nh\ $\gtrsim 7 \times 10^{22}$\cmsq\ at $z\simeq 1.9$).
A DLA could provide such a column if it was dominated by H$_2$ ({Schaye} 2001)
but nonetheless had a low dust-to-gas ratio, which is rather contrived,
or if had a metallicity $\sim$10 times solar, which is rather unlikely.
We conclude that this quasar experiences some intrinsic X-ray absorption and
some from the intervening system, and similarly for its dust reddening.  (There
is moderately strong narrow \ion{C}{4} absorption 7000\,\kms\ from
the quasar redshift, which could be a tracer of high-ionization gas intrinsic 
to the quasar.)  We discuss this issue further in \S\ref{1310} and \S\ref{1323}.
In any case, the combination of moderate dust reddening and strong X-ray 
absorption means that the overall dust-to-gas ratio along the line of sight 
to this quasar is rather small.

\subsubsection{J1226$-$0011; $z_{\rm em}=1.176$, $\Delta(g-i)=0.261$, $\alpha_{\rm UV}=-0.72$, $\Gamma_{\rm HR}=2.01$, $\Delta\alpha_{\rm ox}=0.01$} \label{1226}

The optical spectrum is only mildly red, but 
dust extinction might be present since \delur\ $>$ \delgi\ $>$ \delrz.
There is no indication of X-ray absorption. 
This quasar could be either intrinsically red or mildly dust reddened.

\subsubsection{J1251+0002; $z_{\rm em}=0.878$, $\Delta(g-i)=0.319$, $\alpha_{\rm UV}=-1.01$, $\Gamma_{\rm HR}=1.24$, $\Delta\alpha_{\rm ox}=-0.25$} \label{1251}

The spectrum of this quasar has little rest-frame UV coverage, 
but the evidence of moderate curvature in the spectrum and photometry
[\delur\ $>$ \delgi\ $>$ \delrz] are consistent with dust reddening.
The values of \daox\ and \GHR\ imply significant X-ray absorption.
We conclude that this quasar is dust-reddened with X-ray absorption.

\subsubsection{J1302+0000; $z_{\rm em}=1.797$, $\Delta(g-i)=0.363$, $\alpha_{\rm UV}=-0.86$, $\Gamma_{\rm HR}=1.52$, $\Delta\alpha_{\rm ox}=-0.16$} \label{1302}

The optical spectrum is normal.  While there is intervening absorption, 
there is no indication of spectral curvature from reddening.
The values of \daox\ and \GHR\ are 
consistent with moderate X-ray absorption, but are only $\sim 2\sigma$
from the typical values seen in quasars without X-ray absorption.
We argue that the redness of this object is most likely
intrinsic and not due to dust.

\subsubsection{J1310+0108; $z_{\rm em}=1.392$, $\Delta(g-i)=0.626$, $\alpha_{\rm UV}=-1.61$, $\Gamma_{\rm HR}=0.84$, $\Delta\alpha_{\rm ox}=-0.42$} \label{1310}

The optical spectrum of this quasar is heavily reddened.  There is
strong intervening absorption (a probable DLA) at $z=0.8621$ and
\ion{Ca}{2} absorption, and according to ({Wild} \& {Hewett} (2005) the quasar is
reddened by $E(B-V)=0.209$ assuming an LMC extinction curve.  
However, this quasar is {\em significantly} absorbed in the X-ray, 
consistent with an equivalent solar-metallicity absorbing column of
\nh\,$\simeq (5\pm2) \times 10^{22}$ {\rm ~cm}$^{-2}$.
This is the same situation as for J1133+0058,
and we reach the same conclusion: this quasar 
has both intrinsic and intervening X-ray absorption,
and similarly for its dust reddening.
Unfortunately, the quasar is at too low a redshift to look for intrinsic 
\ion{C}{4} absorption as a possible tracer of high-ionization gas.
We discuss this issue further in \S\ref{1323}.

\subsubsection{J1323$-$0021; $z_{\rm em}=1.388$, $\Delta(g-i)=0.463$, $\alpha_{\rm UV}=-1.12$, $\Gamma_{\rm HR}=1.63$, $\Delta\alpha_{\rm ox}=0.05$} \label{1323}

The optical spectrum of this quasar is heavily reddened, has strong intervening
absorption estimated to be of DLA strength (see \S\ref{sec:dla}), and has
\ion{Ca}{2} absorption that could be accompanied by optical reddening.  (This
intervening system is at too low a redshift to be in the ({Wild} \& {Hewett} (2005) sample.)
This quasar has unremarkable \GHR\ and \daox\ values, but X-ray
spectral-fitting yields $\Gamma=2.1\pm0.4$ and a significant detection
of absorption.  If the X-ray absorption is set to the DLA redshift of
$z=0.716$, then \nh\,$=5.7^{+3.7}_{-3.1}\times10^{21}$\cmsq\ (68\% confidence 
for 2 parameters of interest); an intrinsic absorber requires more than
double that column density (see Table~\ref{tab:xspec}).  

Our absorption detection is consistent with the result of {Peroux} {et~al.} (2006) 
that this is the most metal-rich intervening quasar absorption system known,
with [Zn/H]=+0.61$\pm$0.20 (metallicity $4.1^{+2.4}_{-1.5}$ times solar).  
However, even after adjusting our best-fit solar-metallicity X-ray 
column downward by a factor of four, the resulting \nh\ measured in the X-ray
is still a factor of $9^{+8}_{-6}$ larger than this system's neutral \nhi =
$1.62^{+1.01}_{-0.55}\times 10^{20}$\cmsq\ measured in the UV ({Khare} {et~al.} 2004).

We conclude that an intervening absorption system causes the optical reddening
toward this quasar and some of the X-ray absorption toward it, but not all:
we infer that there is also high-ionization, X-ray-absorbing gas
intrinsic to this quasar.  In fact, the spectrum of this quasar shows 
two strong, narrow \ion{C}{4} absorption systems very near the quasar redshift
({Khare} {et~al.} 2004), at least one of which is accompanied by strong, narrow
\ion{N}{5} absorption in an {\em HST} spectrum (GO project 9382; PI: S. Rao).

It is somewhat suspicious to postulate that there are three quasars in our
sample which have both dust reddening and strong intervening absorption systems
but which also require intrinsic X-ray absorption at the quasar redshift 
(this quasar, J1133+0058 and J1310+0108), even if there is possible evidence
for intrinsic high-ionization gas in the first two cases.
Nonetheless, this conclusion appears more likely than the alternative that the
intervening absorbers are as strong as any ever seen along quasar
sightlines {\em and} have metallicities $\sim$10 times solar.  This is
especially true for J1323$-$0021, where we know the absorber's metallicity and 
\nhi\ and there is still a factor of 10 discrepancy with the X-ray column.  
Moreover, given that dusty intrinsic and intervening absorption systems do
exist, both types of systems will be overrepresented in our red quasar sample,
and rare systems where both effects are at work will be even more 
overrepresented.  
In any case, our conclusion can be tested with 
better X-ray data to constrain the X-ray absorbing columns 
(and to ensure that the measured X-ray columns are not systematically high)
and with high-resolution UV spectroscopy to constrain
the \nhi\ and metallicities of the three intervening absorption systems.

\subsubsection{J1708+6154; $z_{\rm em}=1.415$, $\Delta(g-i)=0.259$, $\alpha_{\rm UV}=-1.13$, $\Gamma_{\rm HR}=1.58$, $\Delta\alpha_{\rm ox}=0.13$} \label{1708}

This quasar's optical spectrum is moderately red.  
It has \delur\ $\simeq$ \delgi\ $>$ \delrz.
The X-ray properties are within the normal range in \daox\ and \GHR.  
All the above evidence is consistent with this quasar being
either intrinsically red or only mildly dust-reddened.

\subsubsection{J1714+6119; $z_{\rm em}=1.847$, $\Delta(g-i)=0.426$, $\alpha_{\rm UV}=-0.97$, $\Gamma_{\rm HR}=1.32$, $\Delta\alpha_{\rm ox}=-0.10$} \label{1714}

In this quasar, the optical spectrum is mildly red.  The X-ray
spectrum is fairly hard, but \daox\ within the normal range.
Given that both the \ion{C}{4} and \ion{He}{2} emission lines in this
quasar are strong and narrow, and that it has red colors in \delrz\ and 
\delgi\ but a blue color in \delur\ (due to strong Ly$\alpha$ emission?),
this object would appear to be intrinsically red rather than dust-reddened.  
({Richards} {et~al.} (2003) presented evidence that quasars with
high equivalent width emission lines tend to be redder.)

\subsubsection{J1715+6323; $z_{\rm em}=2.182$, $\Delta(g-i)=0.352$, $\alpha_{\rm UV}=-1.05$, $\Gamma_{\rm HR}=1.75$, $\Delta\alpha_{\rm ox}=0.02$} \label{1715}

The optical spectrum shows mild reddening. 
The object has red \delrz\ and \delgi\ but blue \delur.  
There is no strong evidence for X-ray absorption from spectral fitting.  
This object is most likely to be intrinsically red.

\subsubsection{J1735+5355; $z_{\rm em}=0.956$, $\Delta(g-i)=0.235$, $\alpha_{\rm UV}=-0.25$, $\Gamma_{\rm HR}=1.73$, $\Delta\alpha_{\rm ox}=0.07$} \label{1735}

In this quasar, which has the bluest $\Delta(g-i)$ value in our
sample, the optical spectrum is normal or somewhat blue (note the
quasar's much bluer position in Figure \ref{fig:ghrdelg}b relative to
the rest of our sample).  The \ion{Fe}{2} emission lines are very weak; this
might be affecting the \delgi\ color.  This quasar would appear to
be neither dust reddened or intrinsically red,
but it does have an abnormal relative color which has caused
it to `contaminate' our sample.

\subsubsection{J1738+5837; $z_{\rm em}=1.279$, $\Delta(g-i)=0.320$, $\alpha_{\rm UV}=-0.70$, $\Gamma_{\rm HR}>0.48$, $\Delta\alpha_{\rm ox}=-0.57$} \label{1738}

Our final quasar has an optical spectrum that is mildly reddened.
It also has \delur\ $>$ \delgi\ $>$ \delrz, consistent with dust
reddening.  The \daox\ and \GHR\ values indicate significantly X-ray
absorption. Thus we conclude that this object is dust-reddened as well
as X-ray absorbed.

\subsubsection{Summary of Notes on Individual Quasars} \label{sec:sum}

The quasars in our relatively red quasar sample appear to have a
variety of dominant causes for their red colors:

\noindent{Intrinsically red: 3 
(\S\ref{1302}, \S\ref{1714}, \S\ref{1715})}

\noindent{Dust-reddened:~3-5~(\S\ref{0002},~\S\ref{1133}?,~\S\ref{1251},~\S\ref{1310}?,~\S\ref{1738})}

\noindent{Either: 2 
(\S\ref{1226}, \S\ref{1708})}

\noindent{Reddened~by~intervening~absorber:~1-3~(\S\ref{1133}?,~\S\ref{1310}?,~\S\ref{1323})}

\noindent{Misclassified as red: 1 
(\S\ref{1735})}

\section{Discussion} \label{sec:dis}

The data presented here show that quasars selected to have the reddest
optical colors at a given redshift are not necessarily X-ray absorbed.  
However, the greater range of optical/UV to X-ray flux ratios and
X-ray hardness ratios in quasars that appear optically red shows that
they have a (marginally significant) tendency to be X-ray absorbed
more often than optically blue quasars.  Unfortunately, it does not
appear straightforward to accurately predict {\em which} red quasars 
will be absorbed in the X-ray using the optical/UV colors alone.  

In our small sample of
twelve relatively red quasars, five show some evidence for intrinsic X-ray
absorption based on the combination of both more negative \daox\ values 
and lower \GHR\ values (when we can constrain \GHR) than normal quasars.
It is plausible that at least two effects are at work:
1) Intrinsically red quasars have intrinsically harder X-ray spectra
but normal \daox\ values ({Gallagher} {et~al.} 2005);
2) Dust-reddened quasars have X-ray spectra hardened (though not necessarily
significantly) by absorption in gas accompanying the observed dust, and
\daox\ values decreased by X-ray absorption (for low dust-to-gas ratios).  

Interpreting the distributions of \daox\ and \GHR\ for relatively 
red quasars can be further complicated by correlations of X-ray and 
dust reddening properties of quasars with other quasar properties.  
{Gallagher} {et~al.} (2005) find that quasars with large \CIV\ blueshifts have
X-ray absorption without significant dust reddening, since such quasars
have relatively blue optical/UV spectra on average ({Richards} {et~al.} 2002).
{Reichard} {et~al.} (2003a) noted that the heavily X-ray-absorbed broad
absorption line quasars are dust-reddened but intrinsically blue.

Thus, even with exploratory X-ray data it can be difficult to 
securely distinguish intrinsically red quasars from 
quasars reddened by SMC-like dust.  
To demonstrate this, in Figure \ref{fig:regs03} we replot \GHR\ vs. 
\daox.  We have added an outline of the region in which $z\sim2$ 
broad absorption line (BAL) quasars are found ({Gallagher} {et~al.} 2006).
We have also added points for quasars with \GHR\ and \daox\ readily
available in the literature; namely, the unusual, optically very blue
and luminous, yet X-ray faint quasar PHL\,1811 ({Leighly}, {Halpern}, \& {Jenkins} 2004; {Choi}, {Leighly}, \& {Matsumoto} 2005)
and the putative soft X-ray weak AGN of {Risaliti} {et~al.} (2003).\footnote{As
pointed out by {Brandt}, {Schneider}, \& {Vignali} (2004) and {Strateva} {et~al.} (2005), the bulk of the
{Risaliti} {et~al.} (2003) objects are not actually X-ray weak if the luminosity
dependence of \aox\ is taken into account.}
The two {Risaliti} {et~al.} (2003) objects with the most negative \daox\ also have the
lowest \GHR\ (the quasars in question are HS 0848+1119 and HS 0854+0915, 
discounting the BAL quasar HS 1415+2701 which has only an upper limit on \daox).
Yet even though both fall on the neutral absorption tracks, 
the lack of spectral curvature in their spectra
points to their being intrinsically red rather than dust reddened
(although there is a $\simeq$10\% chance that HS 0854+0915 is
an unrecognized high-ionization BAL quasar).

Similarly, our best estimate is that the two uppermost filled squares
in Figure \ref{fig:ghrdaox} are quasars with intrinsically red SEDs,
but these two objects also lie on the tracks of neutral absorption.
Thus, falling on the neutral absorption track in this diagram 
is no guarantee that neutral absorption is causing 
the X-ray faintness and spectral flatness.
X-ray spectral fitting may help resolve such a degeneracy, 
but only if the spectrum has sufficient counts to be fit.

Possibly the best candidate 
for an intrinsically red quasar in our {\em Chandra} sample
is J0156+0053 from {Gallagher} {et~al.} (2005).  This object has \GHR$\simeq$1,
\daox$\simeq$0, and an optical/UV spectrum with a red continuum and
strong emission lines.  However, that classification could not be
made without reference to the SDSS spectrum because in all our plots
this object appears consistent with dust reddening (specifically,
$E(B-V)\simeq 0.08$ with a dust-to-gas ratio 10\% of the SMC ratio).

Finally, note that the three leftmost
quasars in Figure~\ref{fig:ghrdaox} 
have properties marginally {\em inconsistent}
with X-ray absorption from gas with a dust-to-gas ratio even just
10\% as large as observed in the SMC, as might be expected from 
gas in an intervening galaxy.
Instead, they are most consistent with absorption from
large columns of partially ionized gas with dust-to-gas ratios
$<$1\% of the SMC value, consistent with the expected 
characteristics of gas in the immediate quasar environment.
These X-ray-faint objects do overlap part of the \GHR-\daox\ 
diagram inhabited by $z \sim 2$ BAL quasars (Gallagher et al. 2006),
but lie much closer to the neutral absorption tracks for their 
redshifts than do most BAL quasars.
These X-ray-faint objects are unlikely to be BAL quasars
but could nonetheless have either ionized intrinsic absorption,
or neutral intrinsic absorption with 90--95\% partial covering.

\subsection{Comparison with $J-K>2$ AGN}	\label{sec:jk2}

Though both samples are referred to as red, our sample and the $J-K>2$
AGN studied in the X-ray by {Wilkes} {et~al.} (2005) are substantially different.
The primary differences are that the $J-K>2$ AGN are less 
luminous and at lower redshift.  Both differences make it more difficult 
in $J-K>2$ samples to isolate the nuclear emission in the optical/near-IR
(from the host galaxy's flux) and in the X-ray (from AGN- or
starburst-related soft excesses and a 
larger fraction of hard-band flux from reflection).

For example, substantial reddening of $J-K>2$ AGN has been claimed
based on the low $I$-band luminosities of many $J-K>2$ AGN in {\em
HST} imaging ({Marble} {et~al.} 2003) and low $R$-band luminosities from {\em
CFHT} imaging ({Hutchings} {et~al.} 2003).  
An equally plausible interpretation is that at least some $J-K>2$
objects are low-luminosity AGN in luminous host galaxies with
substantial near-IR emission from hot nuclear dust, which can produce
quite red $J-K$ colors if present in sufficient quantities
({Scoville} {et~al.} 2000; {Rodr{\'{\i}}guez-Ardila} \& {Mazzalay} 2006).  

If correct, this interpretation would also help explain the X-ray
faintness of the four $J-K>2$ AGN in {Wilkes} {et~al.} (2005) which are not 
type 2 AGN: they would in fact have normal \daox.  
(Note that only one of the four has a value of
$\Gamma$(2-10\,keV) harder than the range spanned by 
our quasar sample, but that they lack \daox\ values because 
their intrinsic nuclear $L_{2500}$ are not known.)
To understand the intrinsic nuclear luminosities of $J-K>2$ AGN,
high-resolution near-IR imaging is needed to directly measure 
the relative AGN and host galaxy contributions 
to the integrated $J$ and $K$ band magnitudes.

\begin{figure}
\epsscale{1.0}
\plotone{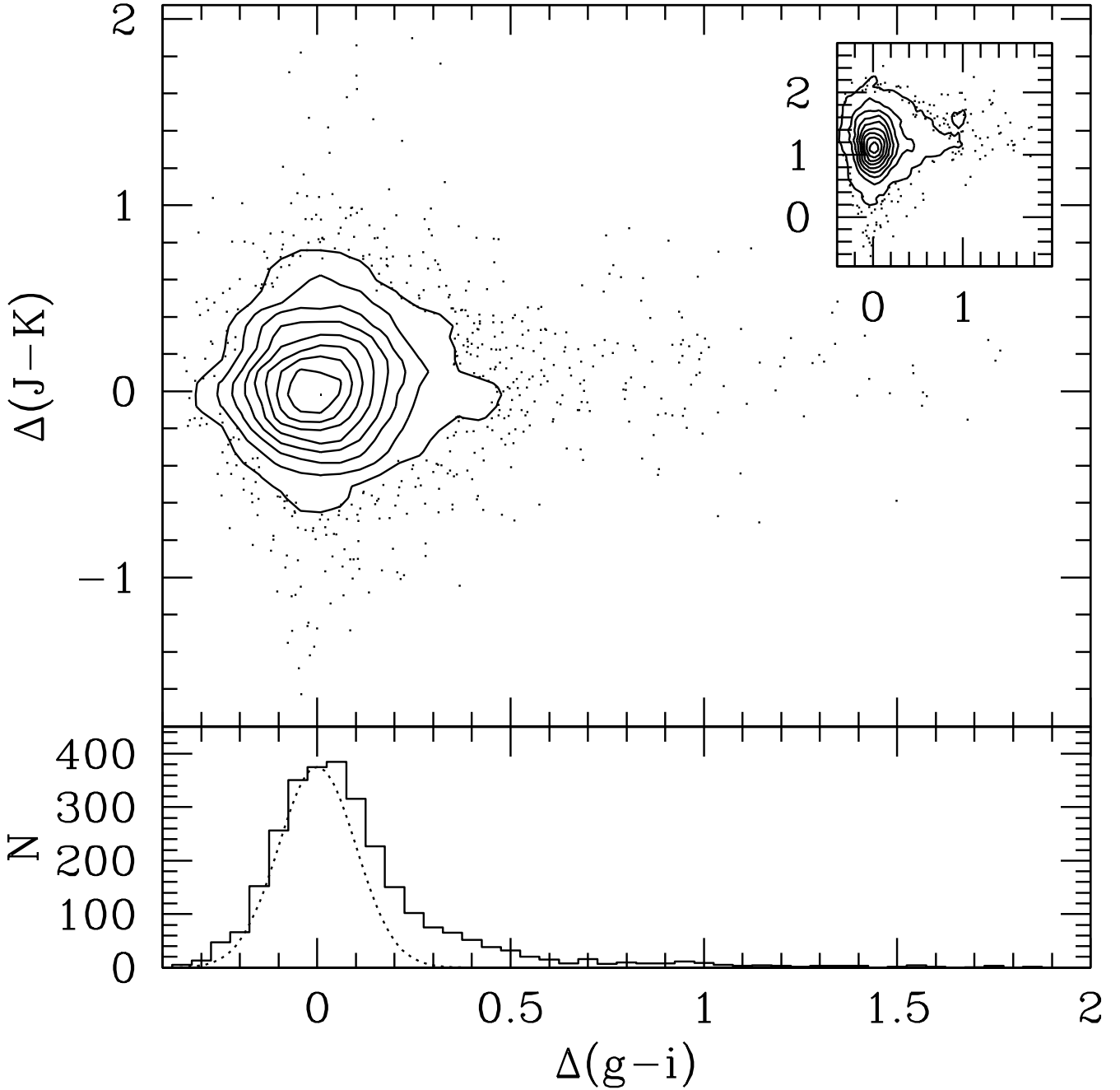}
\caption{A plot of $\Delta(J-K)$ versus \delgi\ for SDSS quasars with 2MASS
photometry.  Both colors have been normalized to the median at each redshift
and so most objects cluster around a value of zero in each normalized color,
as shown by the contours
(points show objects in regions of lower density than the lowest contour level).
The $\Delta(g-i)$ distribution is shown along the bottom, along with
a Gaussian 
overplotted as a dotted line to show a symmetric distribution.
The red (positive) tail of the \delgi\ distribution is symptomatic of
dust reddening ({Richards} {et~al.} 2003), but the $\Delta(J-K)$ distribution
is roughly symmetric and does not show such a tail.
Furthermore, those quasars that are reddest in $\Delta(g-i)$ do not
generally have anomalous $\Delta(J-K)$ colors.
The inset shows $J-K$ vs. \delgi; note that the quasars reddest in 
$\Delta(g-i)$ do not generally meet a $J-K>2$ selection criterion either.
\label{fig:relj-k} }
\end{figure}

While quasar samples selected to have $J-K>2$ do include
red quasars,\footnote{For example, a high polarization subsample 
chosen for spectropolarimetry has revealed some reddened broad-line 
AGN ({Smith} {et~al.} 2003).} the lack of a relation between \delgi\ and
$\Delta(J-K)$ in Figure \ref{fig:relj-k} suggests that
$J-K>2$ selection is not the optimal way to select the most
complete and least contaminated samples of dust-reddened AGN.
Of course, our relatively red quasar sample will miss the most
heavily reddened quasars and has its share of contaminants as well.
Therefore, better samples are needed to understand the dust reddened 
quasar population
and to separate it from the intrinsically red quasar population.
The most promising additional criterion to use in selecting dust-reddened
quasars is spectral curvature, which arises from the extinction properties
of small dust grains.  Spectral curvature should be measurable from
broadband photometry (e.g., \S\ref{0002}), especially if it extends from
the rest-frame far-UV (e.g., {\em GALEX}) through the near-IR,
as well as from well-calibrated, wide-$\lambda$-coverage spectroscopy.

\section{Conclusions}

We have used short \chandra\ observations to demonstrate that
optically red quasars are marginally more likely 
to exhibit evidence for X-ray absorption
than optically blue quasars, but that not all optically red quasars
exhibit X-ray absorption (\S\,\ref{sec:res} and \S\,\ref{sec:dis}).  
Four of the five quasars in our sample with optical spectra most
consistent with dust reddening at the quasar redshift 
(\S\,\ref{sec:sum}) show signatures in \daox\ and \GHR\ of 
X-ray absorption from neutral or partially ionized gas
with column densities \nh\,$\gtrsim 2\times 10^{22} {\rm ~cm}^{-2}$ 
and dust-to-gas ratios $<$1\% of the SMC value 
(\S\,\ref{sec:xo} and \S\,\ref{sec:dis}).
However, half of our sample shows no evidence for such absorption.
We conclude that dust-reddened type 1 AGN (as opposed to fully obscured type 2
AGN) are unlikely to contribute significantly to the remaining unresolved hard
X-ray background, at least for the levels of extinction probed by our sample.

Plausible origins have been identified (see the Appendix)
for most of the intrinsic dispersion exhibited by quasars in \aox\ (the 
correlation of \aox\ with luminosity, plus flux variability in both bands)
and possibly in \GHR\ (the possible correlation of \GHR\ with \auv, plus
X-ray spectral variability).  However, the existence of rare 
outliers like PHL 1811 underscores our ignorance of how those 
parameters interact to produce the observed \aox\ and \GHR\ values.

Progress in understanding the X-ray properties
of the type 1 radio-quiet quasar population will require
even more careful work to create samples that are
as homogeneous as possible (\S\,\ref{sec:jk2}).
Such samples must not only be thoroughly vetted to remove quasars
exhibiting strong intervening absorption and BAL troughs, but should
also span the observed ranges of luminosity, \ion{C}{4} blueshift,
$\alpha_{UV}$, optical/UV spectral curvature, estimated $M_{BH}$,
and possibly emission-line EW, since all those properties may be 
interrelated with a quasar's X-ray properties.
%
%
While such studies can make use of serendipitous archival observations
of random quasars (e.g., {Gallagher} {et~al.} 2005; {Strateva} {et~al.} 2005; {Risaliti} \& {Elvis} 2005), 
the surface density of relatively bright quasars on the sky is low and
the the fraction of the sky covered by hard X-ray observations is small.
Serendipitous observations do not yield samples that cleanly span the
full range of quasar properties, and so there is a continuing need for
targeted X-ray observations of carefully selected quasar samples.

\acknowledgements
This work was enabled by \chandra\ X-ray Center grant
G04--5116X.  PBH acknowledges support from NSERC, SCG from
NASA through the {\em Spitzer} Fellowship Program under award 1256317,  
DMA from the Royal Society, WNB from NASA LTSA grant NAG5--13035,
and DPS from NSF grant AST03--07582.
Funding for the SDSS and SDSS-II has been provided by the Alfred P. Sloan Foundation, the Participating Institutions, the National Science Foundation, the U.S. Department of Energy, the National Aeronautics and Space Administration, the Japanese Monbukagakusho, the Max Planck Society, and the Higher Education Funding Council for England. The SDSS Web Site is http://www.sdss.org/.


\clearpage
\begin{figure}
\epsscale{0.75}
\plotone{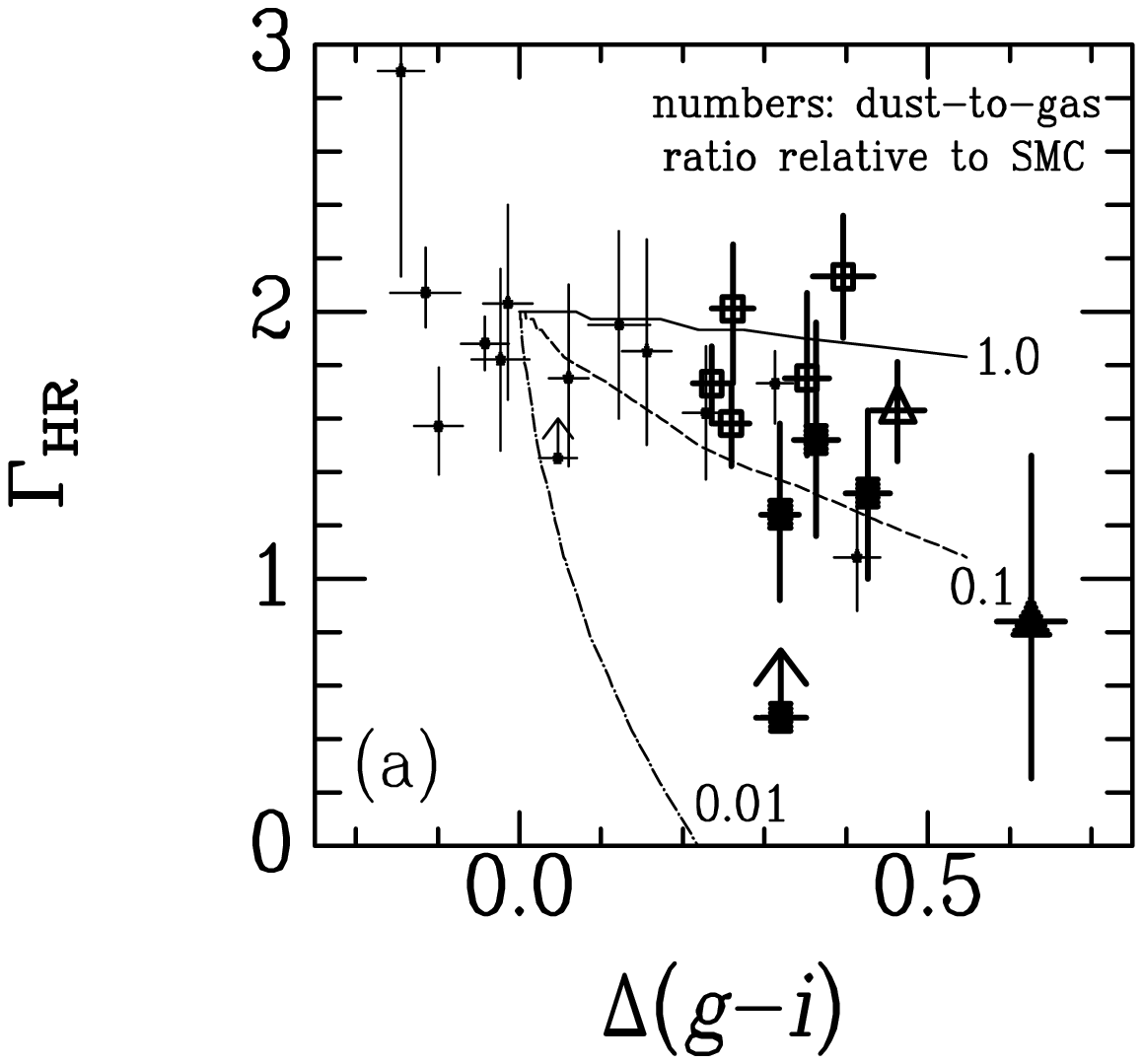}\\ \plotone{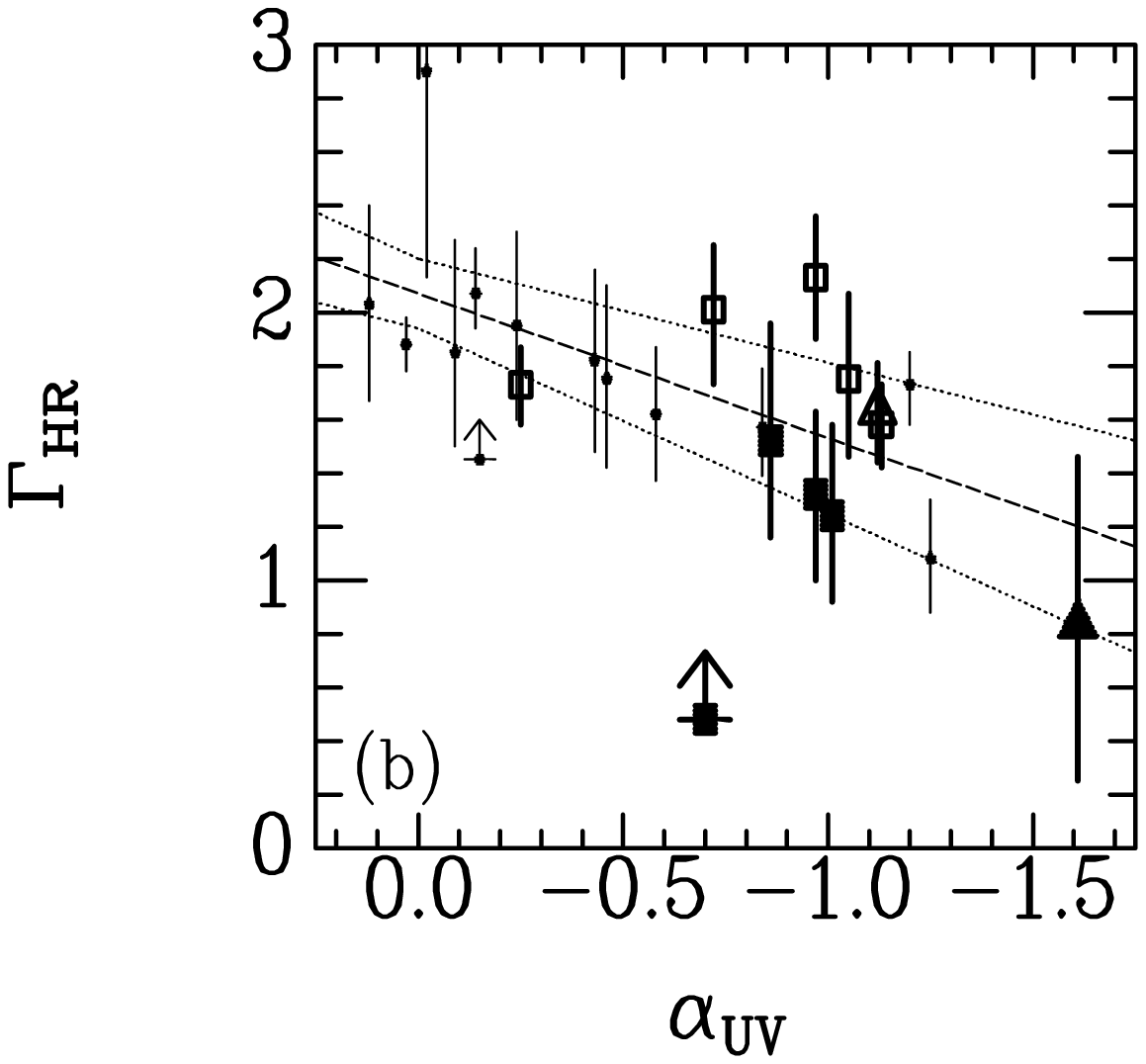}
\caption{X-ray photon index, \GHR, versus two different measures of the
optical/UV color.  Small points are from {Gallagher} {et~al.} (2005) and span the
range of ``normal'' quasars.  Large symbols are our reddened quasar
candidates; open symbols have \daox$>0$.  Triangles indicate quasars
that have probable DLAs along their line of sight.  {\em Panel (a):} Optical
color as determined from the SDSS photometry using \delgi.  Most normal
(unreddened) quasars have $-0.2<$\delgi$<0.2$.  Lines show X-ray
absorption/optical reddening tracks for gas and dust with an SMC-like
extinction curve at $z=1.5$ but with the different labeled dust-to-gas ratios.
The solid track has an SMC dust-to-gas ratio, the dashed track has
a dust-to-gas ratio 0.1 times that of the SMC, and the dot-dashed
track has a dust-to-gas ratio 0.01 times that of the SMC.
The length of each track corresponds to $E(B-V)=0.1$, although the
dot-dashed track only reaches $E(B-V)=0.04$ before going off the plot.
{\em Panel (b):} Optical color as determined from the SDSS spectra by
fitting a power-law to the continuum.  The dashed line is the BCES(Y$|$X)
fit to the full sample, while the dotted lines show the $\pm$1$\sigma$
uncertainties in the fit.
\label{fig:ghrdelg}
}
\end{figure}
\begin{figure}
\epsscale{0.75}
\plotone{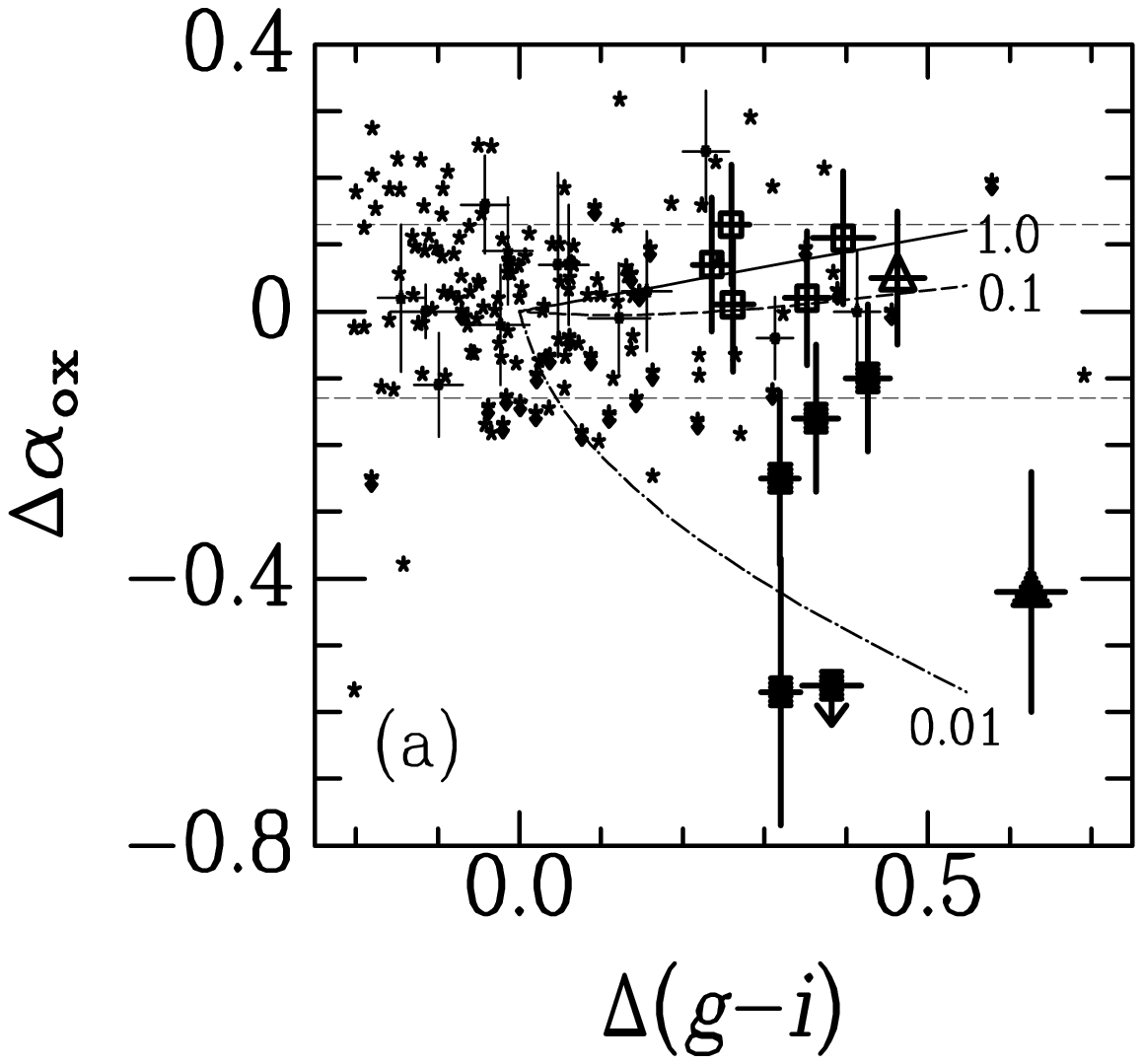}\\ \plotone{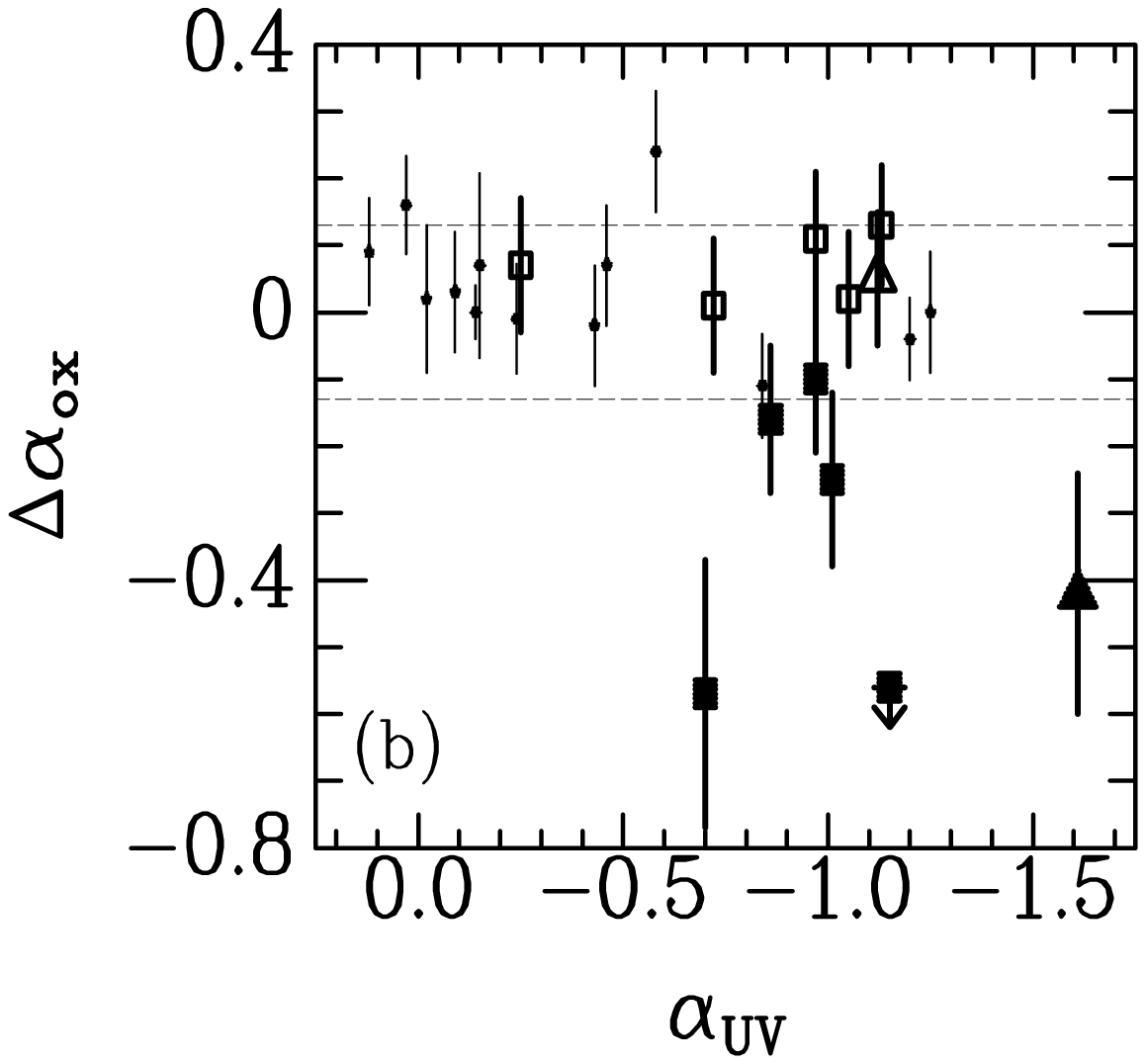}
\caption{Luminosity-corrected optical-to-X-ray flux ratio, \daox,
versus two different measures of the optical color. 
Small points with error bars are from {Gallagher} {et~al.} (2005) and span the
range of ``normal'' quasars.  Large symbols are our reddened quasar
candidates; open symbols have \daox$>0$.  Triangles indicate quasars
that have probable DLAs along their line of sight.
Small stars are from {Strateva} {et~al.} (2005), 
representing SDSS quasars with sensitive {\em ROSAT} observations.
The horizontal dashed lines show the $\pm 1\sigma$ standard 
deviation of the {Strateva} {et~al.} (2005) \daox\ distribution.
{\em Panel (a):} Optical color as determined from the SDSS
photometry using \delgi.  Lines show X-ray absorption/optical reddening
tracks for gas and dust with an SMC-like extinction curve at $z=1.5$ but
with the different labeled dust-to-gas ratios.
The solid track has an SMC dust-to-gas ratio, the dashed track has
a dust-to-gas ratio 0.1 times that of the SMC, and the dot-dashed
track has a dust-to-gas ratio 0.01 times that of the SMC.
The length of each track corresponds to $E(B-V)=0.1$.
{\em Panel (b):} Optical color as determined from the SDSS
spectra by fitting a power-law to the continuum.  \label{fig:daoxdelg}
}
\end{figure}
\begin{figure}
\epsscale{1.2}
\plotone{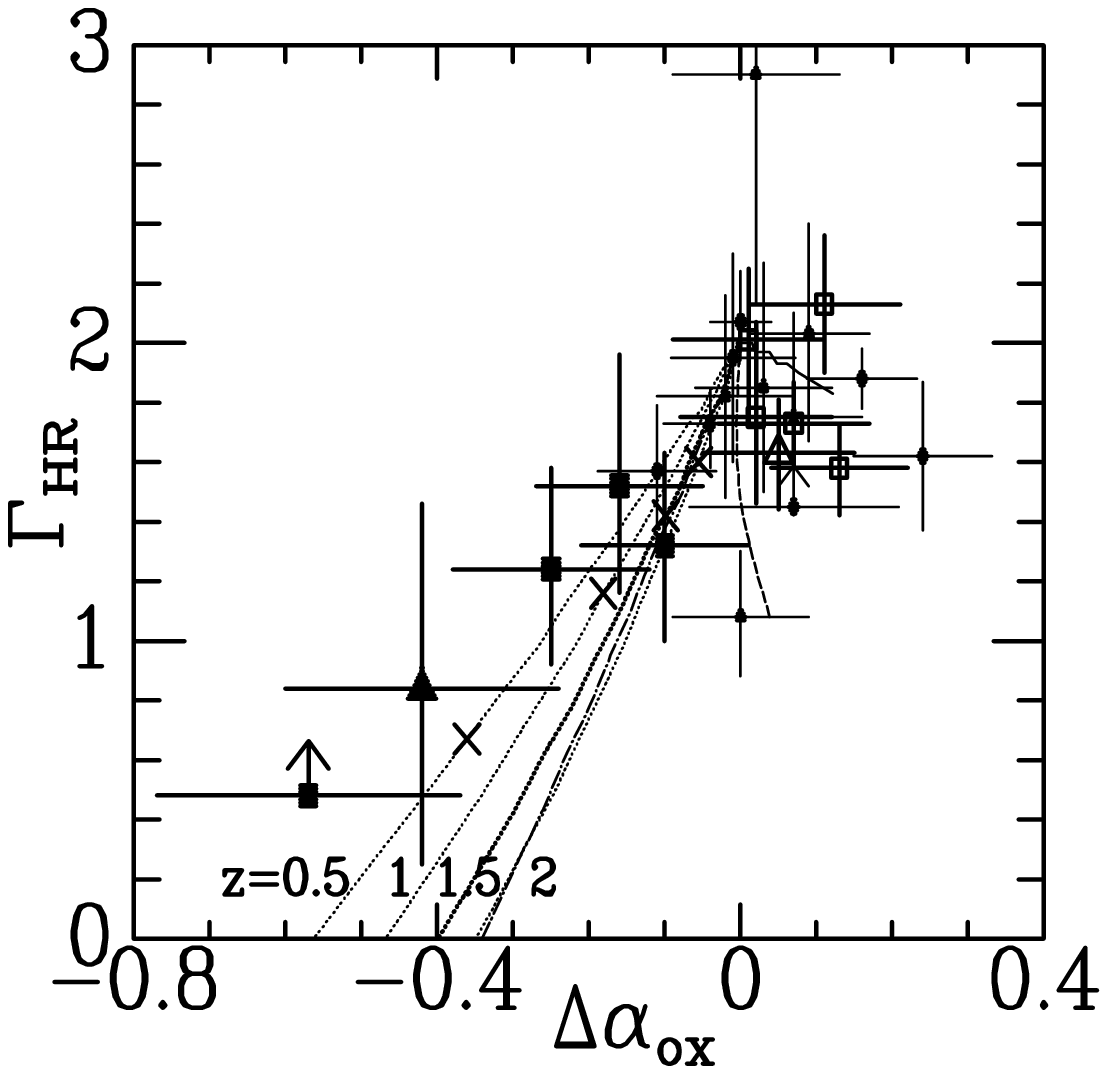}
\caption{
The X-ray spectral index, \GHR, (determined from the observed
hardness ratio) vs.\ \daox, the optical/UV to X-ray index \aox\ relative
to the value expected for the quasar's luminosity.
Small points with error bars are from {Gallagher} {et~al.} (2005) and span the
range of ``normal'' quasars.  Large symbols are our reddened quasar
candidates; open symbols have \daox$>0$.  Triangles indicate quasars
that have probable DLAs along their line of sight.
Dotted lines show the effects of dust-free neutral absorption at the
redshifts indicated.  The crosses on those lines mark the locations of
column densities of \nh\,$=2.5\times 10^{22}$\cmsq\ at each redshift.
The non-dotted lines show X-ray
absorption/optical reddening tracks for gas with SMC-like dust at $z=1.5$.
The solid track has an SMC dust-to-gas ratio, the dashed track has
a dust-to-gas ratio 0.1 times that of the SMC, and the dot-dashed
track has a dust-to-gas ratio 0.01 times that of the SMC.
The length of each track corresponds to $E(B-V)=0.1$, although the
dot-dashed track only reaches $E(B-V)=0.04$ before going off the plot.
\label{fig:ghrdaox}
}
\end{figure}
\begin{figure}
\epsscale{1.0}
\plotone{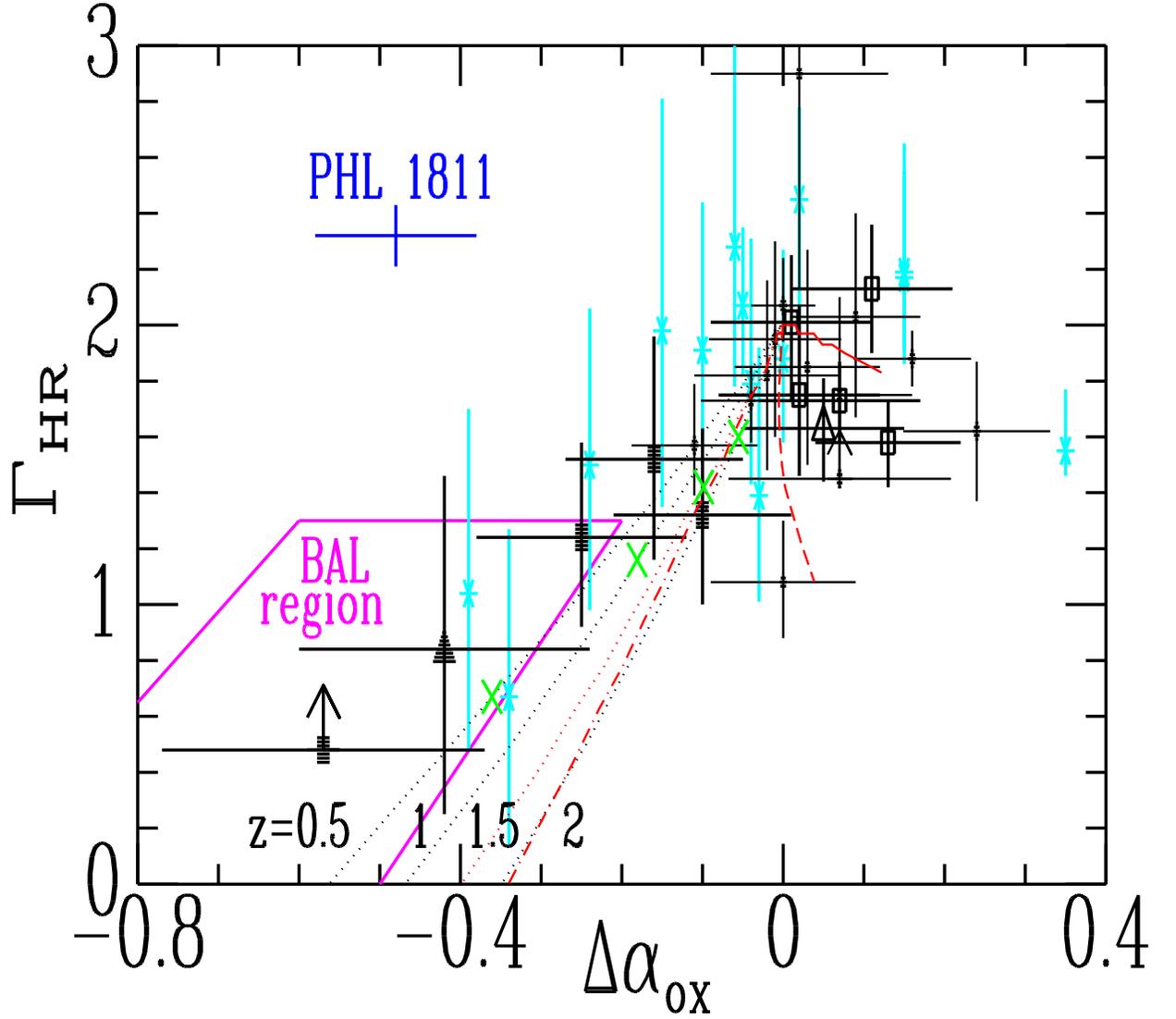}
\caption{The previous Figure with the addition of Risaliti et al.
  (2003) objects (blue asterisks), using numbers taken from their
  Table~1 and assuming an expected \aox=$-$1.73 for their luminosity
  range (their Figure 3).  With that appropriate average \aox\ instead 
  of the value $-1.55$ they used, which is appropriate only for 
  lower-luminosity objects like those found at $z<0.5$ in the 
  Bright Quasar Survey sample ({Brandt} {et~al.} 2000), it
  is clear they have not found a new population of X-ray weak quasars.
  The blue cross marks the position of PHL~1811 ({Choi} {et~al.} 2005), an
  interesting outlier in this parameter space.  The magenta polygon
  outlines the location of the bulk of the BAL~quasars from the sample
  of {Gallagher} {et~al.} (2006).  See \S\ref{sec:dis} for further discussion of
  this Figure.
\label{fig:regs03}
}
\end{figure}
\clearpage
\begin{deluxetable}{lrrrrrrrrcl}
\tabletypesize{\scriptsize}
\rotate
\tablewidth{540pt}
\tablecaption{Target List\label{tab:opt}}
\tablehead{
\colhead{Name (SDSS J)} &
\colhead{$z$}&
\colhead{$i$} &
\colhead{$M_i$} &
\colhead{$\Delta(u-r)$} &
\colhead{$\Delta(g-i)$} &
\colhead{$\Delta(r-z)$} &
\colhead{\auv\tablenotemark{a}} &
\colhead{\nh\tablenotemark{b}} &
\colhead{Spectrum\tablenotemark{c}} &
\colhead{Comments\tablenotemark{d}}
}
\startdata
000230.71$+$004959.0 & 1.355 &  17.91 & $-$26.54  & 0.604$\pm$0.045 & 0.396$\pm0.037$& 0.114$\pm$0.035 & $-0.97$   & 3.14 & 387-575-51791 & dust; DLA?\\
113345.62$+$005813.4 & 1.937 &  17.86 & $-$27.41  & 0.458$\pm$0.071 & 0.382$\pm0.036$& 0.050$\pm$0.065 & $-1.15$   & 2.80 & 282-408-51658 & dust/int?; DLA?, XRF\\
122652.01$-$001159.5 & 1.176 &  17.68 & $-$26.43  & 0.459$\pm$0.026 & 0.261$\pm0.026$& 0.201$\pm$0.031 & $-0.72$   & 1.91 & 289-198-51990 & dust/SED?\\
125140.32$+$000210.8 & 0.878 &  17.43 & $-$25.97  & 0.402$\pm$0.027 & 0.319$\pm0.023$& 0.185$\pm$0.030 & $-1.01$   & 1.59 & 292-559-51609 & dust; XRF\\
130211.04$+$000004.5 & 1.797 &  17.96 & $-$27.16  & 0.191$\pm$0.028 & 0.363$\pm0.027$& 0.193$\pm$0.033 & $-0.86$   & 1.56 & 293-080-51994 & SED; DLA?, XRF\\
131058.13$+$010822.2 & 1.392 &  17.80 & $-$26.66  & 0.741$\pm$0.048 & 0.626$\pm0.042$& 0.444$\pm$0.029 & $-1.61$   & 1.95 & 295-325-51985 & dust/int?; DLA, XRF\\
132323.78$-$002155.2 & 1.388 &  17.61 & $-$26.90  & 0.762$\pm$0.028 & 0.463$\pm0.033$& 0.275$\pm$0.026 & $-1.12$   & 1.89 & 297-267-51959 & int; DLA\\
170817.85$+$615448.5 & 1.415 &  17.64 & $-$26.91  & 0.290$\pm$0.031 & 0.259$\pm0.022$& 0.108$\pm$0.032 & $-1.13$   & 2.51 & 351-550-51780 & dust/SED?\\
171419.23$+$611944.7 & 1.847 &  17.92 & $-$27.27  & 0.089$\pm$0.038 & 0.426$\pm0.027$& 0.334$\pm$0.042 & $-0.97$   & 2.64 & 354-360-51792 & SED\\
171535.96$+$622336.0 & 2.182 &  17.96 & $-$27.59  & 0.116$\pm$0.040 & 0.352$\pm0.027$& 0.296$\pm$0.036 & $-1.05$   & 2.62 & 352-499-51789 & SED\\
173551.92$+$535515.7 & 0.956 &  17.77 & $-$25.84  & 0.228$\pm$0.023 & 0.235$\pm0.023$& 0.073$\pm$0.026 & $-0.25$   & 3.42 & 360-219-51816 & normal\\
173836.16$+$583748.5 & 1.279 &  17.51 & $-$26.81  & 0.377$\pm$0.028 & 0.320$\pm0.024$& 0.185$\pm$0.028 & $-0.70$   & 3.63 & 366-038-52017 & dust; XRF\\
\enddata
\tablenotetext{a}{The spectral index for the ultraviolet continuum (where 
$f_{\nu}\propto\nu^{\alpha_{\rm UV}}$) is measured
by fitting the SDSS spectra.} 
\tablenotetext{b}{The values for \nh\ (in units of 10$^{20}$\cmsq) used in the 
X-ray simulations are from Galactic \HI\ maps ({Dickey} \& {Lockman} 1990).}
\tablenotetext{c}{The SDSS spectrum of each target is identified by the code
{\em plate-fiber-MJD}, giving its spectroscopic {\em plate} number, its {\em fiber} 
number within that plate, and the {\em MJD} on which the plate was observed.}
\tablenotetext{d}{The first comment is our best explanation
for the object's redness (see \S\,\ref{sec:notes}):
{\em dust} -- dust reddening at the quasar redshift;
{\em SED} -- an intrinsically red SED;
{\em int} -- dust reddening by an intervening absorber;
{\em normal} -- apparently misclassified as red.
Other comments: DLA?/DLA -- The SDSS spectrum shows weak/strong evidence for a damped
Ly$\alpha$ absorber candidate;  XRF -- X-ray faint source with \aox\ more than 
1$\sigma$ from the \aox-$L_{2500}$ relation of {Strateva} {et~al.} (2005),
where $\sigma$ is the RMS scatter in \aox\ from Table 5 of {Steffen} {et~al.} (2006).
}
\end{deluxetable}

\begin{deluxetable}{lcccrrrr}
\tabletypesize{\scriptsize}
\tablewidth{0pt}
\tablecaption{{\em Chandra} Observing Log
\label{tab:log}
}
\tablehead{
\colhead{Name} &
\colhead{Obs. ID} &
\colhead{Date (MJD)} &
\colhead{Exposure} &
\multicolumn{2}{c}{Counts\tablenotemark{a}} &
\colhead{Count Rate\tablenotemark{a}} &
\colhead{\HR\tablenotemark{b}} \\
\colhead{(SDSS J)} &
\colhead{} &
\colhead{} &
\colhead{Time (ks)} &
\colhead{Soft} &
\colhead{Hard} &
\colhead{(10$^{-3}$~ct~\persec)} &
\colhead{} 
}
\startdata
000230.71$+$004959.0 &  4861   & 2003 Nov 09 (52952) & 5.67 &   130$^{+12.4}_{-11.4}$ &   20$^{+5.6 }_{ -4.4}$ & 26.63$^{+2.35}_{-2.16}$ & $ -0.73^{+0.07}_{-0.06}$\\
113345.62$+$005813.4 &  4858   & 2004 Mar 29 (53093) & 4.26 &           $<3.8   $     &          $<2.3   $     &          $<0.88  $      &    $\cdots$         \\
122652.01$-$001159.5 &  4865   & 2004 Feb 12 (53047) & 4.90 &    87$^{+10.4}_{ -9.3}$ &   16$^{+5.1 }_{ -4.0}$ & 21.03$^{+2.28}_{-2.07}$ & $ -0.69^{+0.09}_{-0.07}$\\
125140.32$+$000210.8 &  4859   & 2004 Mar 08 (53072) & 4.04 &    27$^{+6.3 }_{ -5.2}$ &   12$^{+4.6 }_{ -3.4}$ &  9.89$^{+1.83}_{-1.56}$ & $ -0.38^{+0.17}_{-0.16}$\\
130211.04$+$000004.5 &  4862   & 2004 May 12 (53150) & 4.90 &    25$^{+6.1 }_{ -5.0}$ &    8$^{+4.0 }_{ -2.8}$ &  6.73$^{+1.39}_{-1.17}$ & $ -0.52^{+0.18}_{-0.16}$\\
131058.13$+$010822.2 &  4854   & 2004 Mar 26 (53090) & 5.45 &     7$^{+3.8 }_{ -2.6}$ &    5$^{+3.4 }_{ -2.2}$ &  2.39$^{+0.86}_{-0.65}$ & $ -0.17^{+0.34}_{-0.32}$\\
132323.78$-$002155.2 &  4855   & 2004 May 09 (53147) & 4.36 &   102$^{+11.1}_{-10.1}$ &   29$^{+6.5 }_{ -5.4}$ & 30.05$^{+2.86}_{-2.62}$ & $ -0.56^{+0.08}_{-0.07}$\\
170817.85$+$615448.5 &  4864   & 2003 Nov 16 (52959) & 4.06 &   153$^{+13.4}_{-12.4}$ &   46$^{+7.8 }_{ -6.8}$ & 49.02$^{+3.73}_{-3.47}$ & $ -0.54^{+0.07}_{-0.06}$\\
171419.23$+$611944.7 &  4856   & 2003 Nov 16 (52959) & 4.62 &    32$^{+6.7 }_{ -5.6}$ &   13$^{+4.7 }_{ -3.6}$ &  9.74$^{+1.68}_{-1.45}$ & $ -0.42^{+0.16}_{-0.14}$\\
171535.96$+$632336.0 &  4857   & 2004 Sep 29 (53277) & 4.25 &    50$^{+8.1 }_{ -7.0}$ &   12$^{+4.6 }_{ -3.4}$ & 14.81$^{+2.11}_{-1.86}$ & $ -0.61^{+0.12}_{-0.10}$\\
173551.92$+$535515.7 &  4863   & 2003 Nov 09 (52952) & 5.42 &   188$^{+14.7}_{-13.7}$ &   47$^{+7.9 }_{ -6.8}$ & 43.34$^{+3.02}_{-2.83}$ & $ -0.60^{+0.06}_{-0.05}$\\
173836.16$+$583748.5 &  4860   & 2003 Nov 16 (52959) & 3.87 &     5$^{+3.4 }_{ -2.2}$ &          $<5.3   $     &  1.81$^{+0.98}_{-0.67}$ &          $<0.03  $     \\
\enddata
\tablenotetext{a}{Detections for the full, soft, and hard bands are
  determined by {\em wavdetect} (as described in \S\ref{sec:data}). Errors are 1$\sigma$ Poisson errors \citep{Gehrels}, except for
  non-detections where the limits are the $90\%$ confidence limits
  from Bayesian statistics ({Kraft}, {Burrows}, \& {Nousek} 1991). The count rate is for the full band, 0.5--8.0~keV.}
\tablenotetext{b}{The \HR\ is defined as $(h-s)/(h+s)$, where $h$ and
  $s$ are the counts in the hard (2.0--8.0~keV) and soft (0.5--2.0~keV)
  bands, respectively. Errors in the \HR\ are propagated from the
  counting errors using the numerical method of {Lyons} (1991).}
 \end{deluxetable}

\begin{deluxetable}{lcccccccr}
\tabletypesize{\scriptsize}
\rotate
\tablewidth{0pt}
\tablecaption{Derived Properties
\label{tab:xcalc}
}
\tablehead{
\colhead{Name (SDSS J)} &
\colhead{\GHR\tablenotemark{a}} &
\colhead{$\log(F_{\rm X})$\tablenotemark{b}} &
\colhead{$\log(f_{\rm 2 keV})$\tablenotemark{c}} &
\colhead{$\log(L_{\rm 2 keV})$\tablenotemark{c}} &
\colhead{$\log(f_{\rm 2500})$\tablenotemark{c}} &
\colhead{$\log(L_{\rm 2500})$\tablenotemark{d}} &
\colhead{\aox} &
\colhead{\daox\tablenotemark{e}} 
}
\startdata
000230.71$+$004959.0 & 2.13$^{+0.23}_{-0.23}$ & $-12.868\pm0.037$ & $-30.577\pm0.037$ &  27.096   & $-26.691$ &  30.983 & $-1.49\pm0.10$  & $ 0.11$\\
113345.62$+$005813.4 &   $\cdots$             & $<-14.029$        & $<-32.287$        & $<25.673$ & $-26.507$ &  31.452 & $<-2.22$        & $<-0.56$\\
122652.01$-$001159.5 & 2.01$^{+0.24}_{-0.28}$ & $-12.942\pm0.045$ & $-30.729\pm0.045$ &  26.826   & $-26.514$ &  31.042 & $-1.62\pm0.10$  & $ 0.01$\\
125140.32$+$000210.8 & 1.24$^{+0.34}_{-0.32}$ & $-13.059\pm0.074$ & $-31.199\pm0.074$ &  26.107   & $-26.445$ &  30.861 & $-1.83\pm0.13$  & $-0.25$\\
130211.04$+$000004.5 & 1.52$^{+0.44}_{-0.36}$ & $-13.309\pm0.083$ & $-31.233\pm0.083$ &  26.668   & $-26.536$ &  31.365 & $-1.81\pm0.11$  & $-0.16$\\
131058.13$+$010822.2 & 0.84$^{+0.62}_{-0.59}$ & $-13.552\pm0.136$ & $-31.913\pm0.136$ &  25.782   & $-26.626$ &  31.069 & $-2.03\pm0.18$  & $-0.42$\\
132323.78$-$002155.2 & 1.63$^{+0.18}_{-0.19}$ & $-12.688\pm0.040$ & $-30.599\pm0.040$ &  27.094   & $-26.514$ &  31.179 & $-1.57\pm0.10$  & $ 0.05$\\
170817.85$+$615448.5 & 1.58$^{+0.15}_{-0.16}$ & $-12.468\pm0.032$ & $-30.400\pm0.032$ &  27.309   & $-26.484$ &  31.225 & $-1.50\pm0.09$  & $ 0.13$\\
171419.23$+$611944.7 & 1.32$^{+0.31}_{-0.32}$ & $-13.087\pm0.070$ & $-31.131\pm0.070$ &  26.791   & $-26.609$ &  31.314 & $-1.74\pm0.11$  & $-0.10$\\
171535.96$+$632336.0 & 1.75$^{+0.32}_{-0.29}$ & $-13.032\pm0.058$ & $-30.791\pm0.058$ &  27.260   & $-26.507$ &  31.545 & $-1.65\pm0.10$  & $ 0.02$\\
173551.92$+$535515.7 & 1.73$^{+0.14}_{-0.15}$ & $-12.557\pm0.029$ & $-30.483\pm0.029$ &  26.896   & $-26.533$ &  30.846 & $-1.51\pm0.10$  & $ 0.07$\\
173836.16$+$583748.5 &         $>0.48$        & $-13.581\pm0.195$ & $-32.176\pm0.195$ &  25.450   & $-26.428$ &  31.197 & $-2.20\pm0.20$  & $-0.57$\\
\enddata
\tablenotetext{a}{\GHR\ is a coarse measure of the hardness of the X-ray
  spectrum determined by comparing the observed \HR\ (see
  Table~\ref{tab:log}) to a simulated \HR\ that takes into account
  temporal variations in the instrument response and the Galactic \nh\
  toward the target (see $\S$\ref{sec:data}).}
\tablenotetext{b}{The full-band X-ray flux, $F_{\rm X}$, has units of \flux\ 
  and is calculated by integrating the power-law spectrum given by \GHR\
  and normalized by the full-band count rate from 0.5--8.0~keV.  The errors
  are derived from the 1$\sigma$ errors in the full-band count rate.}
\tablenotetext{c}{X-ray and optical flux densities were measured at
  rest-frame 2~keV and 2500\,\AA, respectively; units are \fnu.}
\tablenotetext{d}{The rest-frame 2~keV and 2500\,\AA\ luminosity densities
  ($L_{\rm 2 keV}$ and $L_{\rm 2500}$, respectively) have units of \lumin~Hz$^{-1}$.}
\tablenotetext{e}{\daox\ is the difference between the observed \aox\ and the 
expected \aox\ calculated from Equation 6 of {Strateva} {et~al.} (2005) using 
$L_{\rm 2500}$.  The general quasar population has an observed RMS scatter 
around \daox=0 of $\pm$0.146 at $30<\log(L_{2500})<31$ 
(Table 5 of {Steffen} {et~al.} 2006).
}
\end{deluxetable}

\begin{deluxetable}{lcccc}
\tabletypesize{\small}
\tablewidth{0pt}
\tablecaption{Results from X-ray Spectral Fitting\tablenotemark{a}
\label{tab:xspec}
}
\tablehead{
\colhead{Name} &
\colhead{$\Gamma$} &
\colhead{${N}_{\rm H}$} &
\colhead{$C$-stat/$\nu$} &
\colhead{Total} \\
\colhead{(SDSS J)} &
\colhead{} &
\colhead{(10$^{22}$\cmsq)} &
\colhead{} &
\colhead{Counts} 
}
\startdata
000230.71$+$004959.0  & $2.21^{+0.43}_{-0.37}$  & $0.39^{+0.61}_{-0.39}$ & 214/512 & 151\\  
122652.01$-$001159.5  & $2.15^{+0.29}_{-0.26}$  &   $<0.14$              & 210/512 & 103\\
132323.78$-$002155.2  & $2.07^{+0.42}_{-0.40}$  & $1.30^{+0.87}_{-0.78}$ & 230/512 & 131\\

170817.85$+$615448.5  & $1.81^{+0.30}_{-0.30}$  &  $0.46^{+0.54}_{-0.46}$ & 280/51 & 197\\
171535.96$+$632336.0 & $1.67^{+0.41}_{-0.32}$   & $<1.23$               &   183/51 & \phantom{0}62\\
173551.92$+$535515.7 & $1.73^{+0.21}_{-0.17}$  & $<0.19$                &   270/51 & 232 \\
\enddata
\tablenotetext{a}{All spectra were fit with a power-law model with
intrinsic absorption at the quasar redshift.  The errors quoted are
1$\sigma$ (68$\%$ confidence; $\Delta C = 2.30$ for two parameters of
interest). The redshift and Galactic ${N}_{\rm H}$ for each quasar are fixed to
the appropriate values (see Table~\ref{tab:opt}).}
\end{deluxetable}

\clearpage
\appendix{
\section{Intrinsic Scatter in \daox\ and \GHR} 
\label{sec:scat}

Between this paper and {Gallagher} {et~al.} (2005), we have a sample of about
two dozen objects with data of sufficient quality and uniformity to
investigate the intrinsic dispersion in \daox\ and \GHR\ among 
radio-quiet quasars at $z_{em}\sim 1.5$.

Accounting for our measurement errors and excluding the three
clearly anomalous objects with the most negative \daox, all of
which we have classified as predominantly dust reddened, we find
the intrinsic dispersion in \daox\ to be $\sigma=0.05\pm0.02$,
assuming a Gaussian distribution.
(The standard deviation in observed \daox\ in the sample of 
{Strateva} {et~al.} (2005) is $\pm$0.13, but they do not estimate 
the intrinsic dispersion.)  This $\sigma$
is equivalent to a dispersion of $\pm$35\% in the X-ray
luminosity at a fixed UV luminosity.  Scatter in the 
line-of-sight dust extinction will contribute to this dispersion,
but its amplitude is consistent with it being entirely due to
X-ray variability ({Manners}, {Almaini}, \& {Lawrence} 2002).  Our uncertainties already
include a statistical correction for UV variability.

As for \GHR, excluding the two lower limits
we find the observed dispersion in \GHR\ to be
$\sigma_{\Gamma}=0.26\pm0.09$, in good agreement with
{George} {et~al.} (2000) and {Mateos} {et~al.} (2005).
The dispersion in \GHR\ around the correlation with
\auv\ is much lower ($\sigma=0.12\pm0.07$).  Nonetheless, it 
still may be larger than the dispersion in $\Gamma$ attributable
to time variability ($\sigma\simeq 0.06\pm0.02$; {Markowitz}, {Edelson}, \&  {Vaughan} 2003).
We agree with {Risaliti} \& {Elvis} (2005) that the remaining dispersion in \GHR\ is not
attributable to differences in obscuration, because in Figure~\ref{fig:ghrdelg}
even the bluest quasars show a scatter in \GHR.  
Nor is it attributable to trends with redshift or luminosity, 
since \GHR\ does not seem to correlate with either ({Tozzi} {et~al.} 2006).

In summary, we suggest that the observed dispersion in \aox\ seen 
in quasars lacking UV absorption is entirely explained by the known
correlation of \aox\ with luminosity, the effects of flux variability,
and a small contribution from dust extinction.
On the other hand, only $\sim$60\% of the observed dispersion in
\GHR\ is explained by the effects of spectral variability plus the observed
correlation of \GHR\ with \auv.  The remaining dispersion may arise because
\GHR\ does correlate with numerous other optical and ultraviolet spectral
properties (e.g., {Laor} {et~al.} 1994; {Wills} {et~al.} 1999).  Those latter studies, among others, have
suggested that a steep \GHR\ corresponds to a large Eddington ratio.
Thus, although the exact mechanisms which determine the observed values and
trends of \aox\ and \GHR\ in quasars may be unclear, it appears that we have
identified the predominant sources of scatter in those relationships.
}

\end{document}